\documentclass[amsmath,amssymb,reprint,floatfix,aps,pra]{revtex4-2}

\usepackage[normalem]{ulem}
\usepackage[utf8]{inputenc}
\usepackage{textcomp}
\usepackage{physics}
\usepackage{graphicx,amsmath,amssymb}
\usepackage[dvipsnames]{xcolor}
\usepackage{upgreek}
\usepackage[colorlinks=true,citecolor=myblue,linkcolor=myred]{hyperref}
\usepackage{bm}

\definecolor{mygrey}{gray}{0.35}
\definecolor{myblue}{rgb}{0.2,0.2,0.8}
\definecolor{myzard}{cmyk}{0,0,0.05,0}
\definecolor{mywhite}{rgb}{1,1,1}
\definecolor{mywhite}{rgb}{1,1,1}
\definecolor{myred}{rgb}{1,0.,0.3}

\def\ee{{\rm e}}
\def\oo{{\rm o}}
\def\eo{{\rm e/o}}

\newcommand{\nn}{\nonumber}
\newcommand{\tx}{\text}

\newcommand{\rhat}{\hat{r}}
\newcommand{\phihat}{\hat{\phi}}
\newcommand{\zhat}{\hat{z}}

\newcommand{\vecE}{\mathbf{E}}
\newcommand{\vecr}{\mathbf{r}}
\newcommand{\vecg}{\mathbf{g}}

\newcommand{\bcalE}{\bm{\mathcal E}}
\newcommand{\bcalH}{\bm{\mathcal H}}
\newcommand{\bcalO}{\bm{\mathcal O}}

\newcommand{\modeHE}[3]{ \bcalE^{{\tx{#1}_{#2}^{(#3)}}} }
\newcommand{\modeHEc}[3]{ \bcalE^{{{#1}_{#2}^{(#3)}}} }

\newcommand{\modeO}[2]{ \bcalO_{#1}^{#2} }

\newcommand{\mkscomment}[1]{{\color{red}{ *** MKS: #1 ***}}}

\begin{document}
\title{Generation of polarization-entangled counter-propagating photons with high orbital angular momentum}

\author{Elisabeth Wagner}
\email{elisabeth.wagner@students.mq.edu.au}
\affiliation{Department of Physics and CeOPP, University of Paderborn, Warburger Straße 100, D-33098 Paderborn, Germany}
\affiliation{School of Mathematical and Physical Sciences, Macquarie University, NSW 2109, Australia}

\author{Miko\l aj K.~Schmidt}
\affiliation{School of Mathematical and Physical Sciences, Macquarie University, NSW 2109, Australia}

\author{Michael J.~Steel}
\affiliation{School of Mathematical and Physical Sciences, Macquarie University, NSW 2109, Australia}

\author{Polina R.~Sharapova}
\email{polina.sharapova@upb.de}
\affiliation{Department of Physics and CeOPP, University of Paderborn, Warburger Straße 100, D-33098 Paderborn, Germany}

\begin{abstract}
Spin and orbital angular momenta of light are attractive resources to harness for encoding, and manipulating information, with applications in various quantum photonic technologies. However, to fully harness that potential, we require robust sources of high-order angular momentum photons exhibiting nonclassical correlations. Here we propose a fiber-based source of polarization-entangled photons in high-order orbital angular momentum (OAM) modes. In our setup the pairs or photons are generated in a cylindrical fiber through a four-wave mixing process, which induces polarization, or spin entanglement. The photons are then converted to modes exhibiting large OAM by the two helical gratings inscribed in the core of the fiber. We present a complete theoretical framework used to consistently describe this process, and demonstrate a robust control over the joint spectral amplitude of the generated photons.
\end{abstract}

\maketitle

\section{Introduction}

Orbital angular momentum (OAM) modes are a promising tool for holography~\cite{Ren}, high-security encryption~\cite{Fang} and the transmission and manipulation~\cite{Erhard,Fickler2, Barriero} of quantum information due to their potentially unlimited basis dimension~\cite{Allen}. They can be generated  using spatial light modulators~\cite{Fickler1}, spiral phase plates~\cite{Ruffato, Schemmel} or even configurations of nonlinear crystals --- the so-called SU(1,1) interferometers~\cite{Beltran}. However, in free space, optical transmission in the basis of OAM encoded quantum information is associated with significant loss and noise due to air turbulence~\cite{Krenn}. To overcome this problem, guided OAM modes in fibers have attracted a lot of interest~\cite{Brunet_review, Mao, helic, Chen1, opt_vort} because of their associated low transmission losses.

The generation of waveguide OAM modes is conceptually simple: in an ordinary cylindrical step-index waveguide the basis modes are the standard hybrid modes of the fiber, which can be represented as a superposition of OAM modes~\cite{Brunet_review, opt_vort, Okamoto}. Nonetheless, in a standard waveguide without any additional modifications,
 the OAM modes most stable to fiber imperfections   are the modes with zero topological charge. With increasing mode order, the modes become more susceptible to scattering into other states. Supporting stable high-order OAM modes in waveguides is a challenging task which requires  smart dispersion engineering. Indeed, by creating a special refractive index profile of the fiber, one can increase the number of guided OAM modes~\cite{brunet, Gregg, Dashti, Ung, Dai}. In this direction, a promising way to create stable OAM modes in fibers is the adoption of a helical aspect to the refractive index of the fiber core~\cite{LFang, Marshall}.  Current technologies suggest several ways to create such structures including  the point-by-point laser-writing technique~\cite{Williams, Chaboyer, CO2, Zhu}, ultraviolet exposure using a phase mask~\cite{UV}, ion implantation~\cite{Fujimaki}, or thermo-diffusion~\cite{Karpov}.

To apply these concepts towards quantum technologies, here we build on the nonclassical correlations of photons generated in nonlinear processes, such as spontaneous parametric-down conversion~\cite{Jedrkiewicz, Brida, HOM} or spontaneous four-wave-mixing (FWM)~\cite{Boyer1, Boyer2, Helt, Chen}, in a medium with non-zero second- or third-order susceptibility, respectively. In particular, as a source we focus on silica fibers, characterized with a non-zero third-order susceptibility and low propagation losses, for generating entangled photons and transmitting them over large distances.~\cite{Gajda, Anjum, Cordier}.

In this work, we theoretically investigate the generation of counterpropagating photons which possess high-order OAM and are entangled in their polarization degree of freedom. The photons are created via a spontaneous FWM process inside a silica fiber with helical gratings.

The manuscript is structured as follows: In Section~\ref{sec:modes} we define the OAM modes of step-index waveguides; in Section~\ref{sec:fwm} we consider the FWM processes in the absence (Subsection~\ref{sec:no.gratings}), and presence of the helical gratings in optical fibers (Subsection~\ref{subsec:helical}), and revisit the asymptotic-in and -out formalism (Subsection~\ref{subsec:asymptotic},\cite{asymptotic}). We then formulate the expressions for the joint spectral intensities of the generated OAM modes, present numerical results and discuss entanglement in Section~\ref{sec:generation}, and summarize our findings in Section~\ref{sec:conclusion}.

\section{Modes of a step-index waveguide}\label{sec:modes}
We begin by carefully describing the notation and field structure of the various waveguide modes involved. The notation is necessarily rather involved, and it is important to notationally distinguish the \emph{label} of a mode, which we write in upright font, from the \emph{fields} comprising the mode.

In general, the guided modes of step-index fibers are described by the hybrid modes
$\tx{HE}_{m,n}^{{(\eo)}}$ and $\tx
{EH}_{m,n}^{{(\eo)}}$.
The superscript (e/o) designates even/odd polarization corresponding to a $90^\circ$  rotation in the spin degree of freedom --- the simplest example is the choice of horizontal/vertical polarization.
The subscripts $m,n$ are non-negative integers that label respectively the azimuthal and radial number of the mode.

The real electric field distribution of the mode $\bf E(\mathbf{r},t)$ is  represented in cylindrical coordinates $(r,\phi,z)$ such that the transverse ($r,\phi$) and longitudinal ($z$) dependency can be  separated as follows:
\begin{align}
    \mathbf{E}(\mathbf{r},t)
    & = \vecE^{(+)}(\vecr,t) + \vecE^{(-)}(\vecr,t) \nn\\
    &= E_0 \; \bcalE(r,\phi) \; e^{i(k_\mu(\omega) z - \omega t)} + \text{c.c.},
\label{eq:E}
\end{align}
where $E_0$ is overall the field amplitude, ultimately determined by the laser power.
Note that the dispersion relation between the frequency $\omega$ and wavevector $k_\mu(\omega)$ is a  characteristic of each particular mode, with $\mu$ standing for all the mode and polarization indices involved.
Since by convention $k_\mu(\omega) > 0 $, Eq.~\eqref{eq:E} corresponds to a mode propagating in the positive direction $+z$, whereas a backwards propagating mode carries the longitudinal phase modulation  $\exp[-ik_\mu(\omega) z -i \omega t]$.

Next, $\bcalE(r,\phi)$ denotes the complex transverse mode profile which defines the type of the mode --- it can be any of the two hybrid (HE and EH) or the transverse electric (TE) and magnetic (TM) mode families.
In  the cylindrical basis $(\rhat,\phihat,\zhat)$,
the mode profile can be written compactly as an element-wise product ($\circ$) of the $r$- and $\phi$-dependent vectors:
\begin{align}
    \bcalE(r,\phi)=\mathbf{e}(r)\circ\mathbf{g}(\phi),
    \label{eq:fieldprofile}
\end{align}
with the full analytical expressions of the radial functions $\mathbf{e}_{m,n}^{\mathrm{HE}/\mathrm{EH}}(r)$ of the hybrid modes given in Eq.~\eqref{eq:radialpart} of Appendix~\ref{sec:waveguides}.
The angular dependence is defined by
\begin{subequations}
\begin{align}
    \vecg_m^{{(\ee)}}(\phi) &= \qty[\cos(m\phi), \sin(m\phi), \cos(m\phi) ]^T,
    \label{eq:ge}\\
    \vecg_m^{{(\oo)}}(\phi) &= \qty[\sin(m\phi), -\cos(m\phi), \sin(m\phi) ]^T,
\end{align}\label{eq:g}\end{subequations}
which implies that even and odd polarized modes  are orthogonal to each other if they exhibit the same azimuthal number $m$. This follows as all azimuthal overlap integrals of the mode components vanish: $\int_0^{2\pi} \tx d\phi \, \qty[\vecg_m^{{(\ee)}}(\phi)]_\mu \cdot \qty[\vecg_m^{{(\oo)}}(\phi)]_\mu=0$ for all $\mu \in \{r,\phi,z\}$.

Combining Eqs.~\eqref{eq:fieldprofile} and~\eqref{eq:g}, the total mode functions $\bcalE(r,\phi)$ of the hybrid modes $\tx{HE/EH}_{m,n}^{\tx{(e/o)}}$ are then:\cite{Okamoto}
\begin{subequations}
\begin{align}
    \modeHEc{\cal M}{m,n}{\ee}(r, \phi)
    &= \rhat\left[\mathbf{e}_{m,n}^{\cal M}(r)\right]_r \cos(m\phi) \nn\\
    &\ \ \ + \phihat\left[\mathbf{e}_{m,n}^{\cal M}(r)\right]_\phi \sin(m\phi)\nn\\
    &\ \ \ + \zhat\left[\mathbf{e}_{m,n}^{\cal M}(r)\right]_z \cos(m\phi),
    \label{eq:evenhybrid}\\
    \modeHEc{{\cal M}}{m,n}{\oo}(r, \phi)
    &= \rhat\left[\mathbf{e}_{m,n}^{\cal M}(r)\right]_r \sin(m\phi) \nn\\
    &\ \ \ - \phihat\left[\mathbf{e}_{m,n}^{\cal M}(r)\right]_\phi \cos(m\phi)\nn\\
    &\ \ \ + \zhat\left[\mathbf{e}_{m,n}^{\cal M}(r)\right]_z \sin(m\phi)
\end{align}
\end{subequations}
where $\cal M$ denotes either of the mode labels HE or EH.

From the above modes we can construct modes with well-defined spin and orbital angular momenta (SAM and OAM), as follows (we refer the reader to Table~\ref{tab:samoamexplain} for several examples):
First, the modes with vanishing OAM charge and (left/right) circular polarization, are given by a superposition of the \textit{fundamental} hybrid modes $\tx{HE}_{1,1}^{(\eo)}$:
\begin{align}
\bcalE^{\Sigma^\pm}(r, \phi)
=\modeHE{HE}{1,1}{\ee}(r, \phi)\pm i \,
\modeHE{HE}{1,1}{\oo}(r, \phi).
\end{align}

Similarly, the OAM modes, or modes with non-zero OAM, can be represented as a superposition of higher-order hybrid modes.
However, as a result of the spin-orbit coupling effect, to formulate the OAM modes one has to consider also the modes' SAM, denoted by $\sigma=\pm$.
Specifically, if the SAM exhibits the same handedness as the OAM of the mode, i.e.~if the angular momenta are co-rotating, the normalized mode
$\mathbf{O}_{\pm m,n}^{\pm}$
is defined by~\cite{opt_vort, Mao, Brunet_review} mode functions $\bcalE(r,\phi)$ of the form:
\begin{align}
    &\modeO{\pm m,n}{\pm}  \nn\\
    &=\frac{1}{\sqrt{2}} \qty(\modeHE{HE}{m+1,n}{\ee} \pm i\ \modeHE{HE}{m+1,n}{\oo} ) \nn \\
    &= \frac{1}{\sqrt{2}} \left\{\rhat\qty[\mathbf{e}_{m+1}^\text{HE}(r)]_r \mp i\phihat\qty[\mathbf{e}_{m+1}^\text{HE}(r)]_\phi + \zhat\qty[\mathbf{e}_{m+1}^\text{HE}(r)]_z \right\} \nn \\
    &\ \ \ \ \ \ \ \ \  \times e^{\pm i (m+1) \phi},
    \label{eq:Oplusdef}
\end{align}
with the upper index labeling the SAM ($\sigma=\pm$), and the first lower index the OAM ($\pm m$ with $m>0$).
 Conversely, for modes $\mathbf{O}_{\pm m,n}^{\mp}$ with counter-rotating angular momenta (opposite handedness of SAM and OAM), we define
\begin{align}
& \modeO{\pm m,n}{\mp}  \nn\\
    &= \frac{1}{\sqrt{2}}\qty(
    \modeHE{EH}{m-1,n}{\ee} \pm i
    \modeHE{EH}{m-1,n}{\oo} )\nn\\ &= \frac{1}{\sqrt{2}}\left\{\rhat\left[\mathbf{e}_{m-1}^\text{EH}(r)\right]_r \mp i\phihat\left[\mathbf{e}_{m-1}^\text{EH}(r)\right]_\phi + \zhat\left[\mathbf{e}_{m-1}^\text{EH}(r)\right]_z \right\} \nn \\
    &\ \ \ \ \ \ \ \ \  \times e^{\pm i(m-1)\phi},
    \label{eq:Ominusdef}
\end{align}
where the OAM order is restricted to $m\geq2$ in the last expression.

For both modes in Eqs.~\eqref{eq:Oplusdef} and~\eqref{eq:Ominusdef}, the azimuthal $(m)$ and radial $(n)$ indices of the HE and EH hybrid modes and the OAM modes are positive integers. The OAM modes with negative azimuthal number, $-m$, exhibit an OAM mode profile with opposite handedness.
Since it is significantly harder to pump the higher order radial modes, we neglect them and set the radial number equal to $n=1$ in further consideration.
Furthermore, all the transverse modes are normalized by the Poynting vector component in the longitudinal direction $\hat{z}$:
\begin{equation}\label{eq:normalization}
    \bcalE(\bf r)\rightarrow\frac{\bcalE(\bf r)}{\sqrt{{\cal P}_z}},
\end{equation}
where
\begin{align}
{\cal P}_z = & \iint \tx d^2 \vecr \,  \mathbf{E}(\vecr) \times \mathbf{H}(\vecr) \\
 = & \iint \tx d^2 \vecr \, \{[\bcalE(\mathbf{r})]^* \times \bcalH(\mathbf{r})+\bcalE(\mathbf{r}) \times [\bcalH(\mathbf{r})]^*\}.
\end{align}
\begin{table}
\begin{tabular}{l|c|c|c|c|c|c|c|c}
Mode & $\left[(\bcalE)_r, (\bcalE)_\phi\right]$ & SAM & OAM \\\hline
$\modeHE{HE}{1,1}{\ee}$ & $\left[\left(\mathbf{e}_1^\text{HE}\right)_r \cos\phi, \left(\mathbf{e}_1^\text{HE}\right)_\phi\sin\phi\right]$ &  0 & 0 \\
$\modeHE{HE}{1,1}{\oo}$ & $\left[\left(\mathbf{e}_1^\text{HE}\right)_r \sin\phi, -\left(\mathbf{e}_1^\text{HE}\right)_\phi\cos\phi\right]$ & 0 & 0 \\
$\modeHE{HE}{0,1}{\ee}\equiv \text{TM}_{0,1} $ & $\left[1,0\right]$ & 0 & 0  \\
$\modeHE{HE}{0,1}{\oo}\equiv \text{TE}_{0,1} $ & $\left[0,-1\right]$ & 0 & 0 \\
$\bcalE^{\Sigma^+}$ & $\left[\left(\mathbf{e}_1^\text{HE}\right)_r, -i\left(\mathbf{e}_1^\text{HE}\right)_\phi\right]e^{i\phi}$ & 1 & 0 \\
$\bcalE^{\Sigma^-}$ & $\left[\left(\mathbf{e}_1^\text{HE}\right)_r,i\left(\mathbf{e}_1^\text{HE}\right)_\phi\right]e^{-i\phi}$ & -1 & 0 \\
$\modeO{1,1}{+}$ & $\left[\left(\mathbf{e}_2^\text{HE}\right)_r , -i\left(\mathbf{e}_2^\text{HE}\right)_\phi \right]e^{2i\phi}$ & 1 & 1 \\
$\modeO{2,1}{+}$ & $\left[\left(\mathbf{e}_3^\text{HE}\right)_r , -i\left(\mathbf{e}_3^\text{HE}\right)_\phi \right]e^{3i\phi}$ & 1 & 2 \\
$\modeO{2,1}{-}$ & $\left[\left(\mathbf{e}_1^\text{HE}\right)_r , -i\left(\mathbf{e}_1^\text{HE}\right)_\phi \right]e^{i\phi}$ & -1 & 2 \\
$\modeO{-1,1}{-}$ & $\left[\left(\mathbf{e}_2^\text{HE}\right)_r , i\left(\mathbf{e}_2^\text{HE}\right)_\phi \right]e^{-2i\phi}$ & -1 & -1 \\
\end{tabular}
\caption{Table of SAM and OAM for a few low-order modes. The second column denotes the azimuthal dependence of the radial and azimuthal components of the electric field, and the final two columns give the spin and orbital angular momentum in units of $\hbar$.}
\label{tab:samoamexplain}
\end{table}

\section{Spontaneous four-wave-mixing in fibers}\label{sec:fwm}

We now turn to the generation of OAM states through spontaneous FWM in a silicon fiber. We first consider the general description of the FWM process in a system without the helical grating (Subsection~\ref{sec:no.gratings})
using both the hybrid as well as the OAM modes as basis states for the generated signal and idler photons.
Then, we discuss how laser-written gratings can effectively couple the optical modes (Subsection~\ref{subsec:helical}), modify the dispersion, and thus increase the efficiency of the OAM mode generation process.

\subsection{Four-wave mixing in fibers without gratings}\label{sec:no.gratings}

\begin{figure}[hbtp]
    \includegraphics[width=\columnwidth]{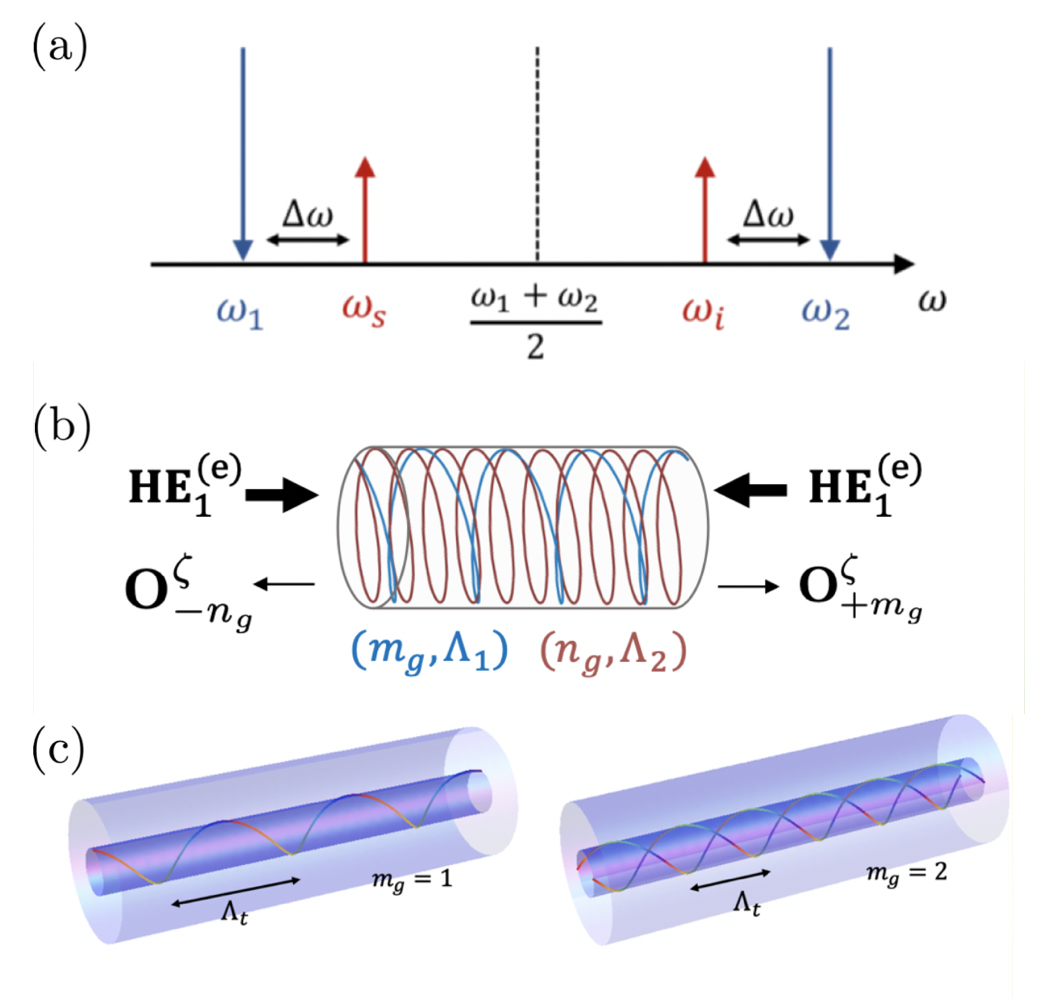}
    \caption{
    (a) Frequency conservation diagram for the considered four-wave-mixing (FWM) process which includes the signal ($\omega_s$), idler ($\omega_i$), and the two pump photons ($\omega_{1,2}$).
    (b) Setup of the spontaneous FWM process in a helical grating fiber. Two continuous-wave (CW) pump beams, both in the fundamental mode
    $\mathbf{HE}_1^{(\ee)}$, counterpropagate through the fiber with two right-handed transmission gratings of different orders $m_g$ and $n_g$, and grating periods $\Lambda_1$ and $\Lambda_2$ respectively. While the two pump beams are far frequency-detuned from the gratings, they generate signal and idler photons tuned to the gratings resonance through the FWM process depicted in (a). The photons emerge from the opposite ends of the fiber in the associated OAM modes $\mathbf{O}_{m_g}^{\zeta}$ and $\mathbf{O}_{-n_g}^{\zeta}$ (see definition of these modes in the text).
    (c) Fibers with  spiral-shaped helical gratings in the cores with periods $\Lambda_\mathrm{t}$ and different topological charges: $m_g=1$ (left) and $m_g=2$ (right).
    }\label{fig:grating}
\end{figure}

The nonlinear FWM process under consideration is depicted in Fig.~\ref{fig:grating}(a): using counterpropagating, strong and coherent pump beams with frequencies $\omega_1$ and $\omega_2$, pairs of counterpropagating signal and idler photons are generated with frequencies $\omega_s$ and $\omega_i$, respectively.
This process is governed by the third-order nonlinear electric susceptibility $\chi^{(3)}$, and is described by the nonlinear Hamiltonian~\cite{boyd}
\begin{align}
	&\hat H_{_{\tx{NL}}}(t) =\varepsilon_0
    \int_V \mathrm{d}^3 \textbf{r}\ \sum_{\mu\nu\tau\rho} \chi_{\mu\nu\tau\rho}^{(3)} \nn\\
    &\times
    \left(\textbf{E}_{1}^{(+)}\right)_\mu \,
    \left(\textbf{E}_{2}^{(+)}\right)_\nu \,
    \left(\hat{\textbf{E}}_{s}^{(-)}\right)_\tau \,
    \left(\hat{\textbf{E}}_{i}^{(-)}\right)_\rho \,
    +\; {\textup{h.c.}},
     \label{eq:H}
\end{align}
where the superscripts $(+)$ and $(-)$ indicate the respective positive or negative frequency parts of each of the electric fields, and the summation includes all three components of the modes with indices $\mu,\nu,\tau,\rho \in \{x,y,z\}$.

Both pump beams are considered to be in the
$\mathbf{HE}_1^{(\ee)}$ mode, propagating in opposite directions as shown in Fig.~\ref{fig:grating}(b).
To describe the generated signal and idler photons, we can embrace any convenient basis of modes of the fibers. Here, we will first show the arguably simpler treatment using the hybrid modes, and later contrast it by using the OAM modes as basis states. The latter should be better suited to describe coupling due to the helical grating in Subsection~\ref{subsec:helical}.

\subsubsection{Formulation of the FWM in hybrid modes}

In a fiber without gratings, the signal and idler photons can be generated  not only  in the fundamental modes, but in various higher-order modes $\mathbf{HE}_{m+1}^{(\eo)}$
and  $\mathbf{EH}_{m'-1}^{(\eo)}$,
where $m\in\mathbb{N}_{\geq 1}$ and $m'\in\mathbb{N}_{\geq 2}$.
The efficiency of such a process is defined by the phase-matching dictated by the longitudinal ($\int \mathrm{d}z(\ldots)$) as well as the transverse ($\iint \mathrm{d}^2\mathbf{r}(\ldots)$) integrals of the nonlinear Hamiltonian in Eq.~\eqref{eq:H}. The transverse integral refers to the modal overlap and can, for the generated HE modes, be written as
\begin{align}
    & I^{m_s,\eo;m_i,\eo}_\tx{HE,HE} =
    \sum_{\mu\nu\tau\rho}\;\chi^{(3)}_{\mu\nu\tau\rho} \iint \mathrm{d}^2\mathbf{r}
          \nn\\
        	 & \times
            \left(\modeHE{HE}{1}{\ee} \right)_{\!\mu}
            \left(\modeHE{HE}{1}{\ee} \right)_{\!\nu}
            \left(\modeHE{HE}{m_s+1}{\eo} \,\right)^{\!*}_{\!\tau}
            \left(\modeHE{HE}{m_i+1}{\eo} \,\right)^{\!*}_{\!\rho},
        \label{eq:It}
\end{align}
using the notation convention of Eq.~\eqref{eq:E}. Note that while omitted for clarity, the overlap integral does implicitly depend on the frequencies of the signal ($\omega_s$) and idler ($\omega_i$) photons via the frequency dependence of the mode profiles.


Further, the signal and idler photons are counterpropagating and can only be generated as pairs of photons with the same polarization, i.e.~$\mathbf{HE}_{m_s+1}^{(\ee)} \mathbf{HE}_{m_i+1}^{(\ee)}$ or $\mathbf{HE}_{m_s+1}^{(\oo)} \mathbf{HE}_{m_i+1}^{(\oo)}$ --- the generation of signal and idler photons with orthogonal polarizations is forbidden due to the zero azimuthal integrals (similar to the argument following Eq.~\eqref{eq:g}).
Moreover, the probabilities of generating the $\mathbf{HE}_{m_s+1}^{(\ee)} \mathbf{HE}_{m_i+1}^{(\ee)}$ and $\mathbf{HE}_{m_s+1}^{(\oo)} \mathbf{HE}_{m_i+1}^{(\oo)}$ pairs coincide since the corresponding azimuthal integrals are equal and the radial profiles of the modes are independent of their (even/odd) polarization, see~Eq.~\eqref{eq:radialpart}.

\subsubsection{Formulation of the FWM in the OAM basis}

Describing the generation of FWM photons in the OAM basis, one can analyze how the FWM process conserves the net angular momentum and the separate OAM and the SAM of the input state.
To this end, we use the expressions for  the co- and counterpropagating OAM mode profiles $\mathbf{O}_{\pm m}^\pm$ and $\mathbf{O}_{\pm m}^\mp$ as defined in Eqs.~\eqref{eq:Oplusdef} and~\eqref{eq:Ominusdef} to define the corresponding transverse overlap integral:
\begin{align}
     &I^{\pm m_s,\sigma_s;\pm m_i,\sigma_i} = \sum_{\mu\nu\tau\rho}\;\chi^{(3)}_{\mu\nu\tau\rho}
        \iint \mathrm{d}^2\mathbf{r} \nn\\
       & \ \ \  \times \left[
        	\left(\modeHE{HE}{1}{\ee}\right)_{\!\mu}
            \left(\modeHE{HE}{1}{\ee}\right)_{\!\nu}
            \left(\modeO{\pm m_s}{\sigma_s}\,\right)_{\!\tau}^{\!*}
            \left(\modeO{\pm m_i}{\sigma_i}\,\right)_{\!\rho}^{\!*}
        \right].
        \label{eq:IOO}
\end{align}
%
Note that the handednesses of the OAM and SAM of the modes are not necessarily correlated, i.e.~the $\pm$ signs of the OAM orders $m_s$ and $m_i$ are independent both of each other and of the associated SAM handednesses $\sigma_{s,i}=\pm$.
The explicit form of 
$I^{\pm m_s,\sigma_s;\pm m_i,\sigma_i}$ is given in Appendix~\ref{sec:overlapintegral} under the assumption of a homogeneous medium with isotropic third-order susceptibility tensor $\chi^{(3)}$.


From  the azimuthal integral in Eq.~\eqref{eq:IOO}, we can determine selection rules for the generated signal and idler OAM modes.
From the requirement  that 
$I^{\pm m_s,\sigma_s;\pm m_i,\sigma_i}$ is non-zero, and in particular so is its azimuthal integral contribution,
we find the angular momentum conservation rule
\begin{align}\label{eq:mcondition}
    \pm m_s \pm m_i + \tilde\sigma_s + \tilde\sigma_i \in \{\pm 2,\,0\},
\end{align}
where, as before, the two $\pm$ signs of the OAM numbers are chosen independently. If the sum of OAM ($\pm m_s \pm m_i$) and SAM (the polarizations of the signal ($\tilde\sigma_s=\pm 1$) and the idler ($\tilde\sigma_i=\pm 1$) photons~\footnote{For accuracy, we use two different variables for describing the polarization of the modes: $\tilde\sigma_{s,i}\in\{+1,-1\}$ labels the total value of the SAM while $\sigma_{s,i}\in\{+,-\}$ represents only their handedness.}) is $\pm2$, then the process exhibits a spin-orbit coupling effect. The latter will not be further discussed in this work because the efficiency of this process is drastically reduced in comparison to the FWM process without spin-orbit coupling as discussed later in Section \ref{sec:3B1}.

A few instances of possible generations of (counterpropagating) OAM modes which would satisfy Eq.~\eqref{eq:mcondition} are listed below:
\begin{subequations}
\begin{align}\label{eq:first}
    \mathbf{O}_{+m}^{+} \tx{ and } \mathbf{O}_{-m}^- &&\tx{ where } &&(+m-m+1-1=0),
\end{align}
\begin{align}
    \mathbf{O}_{+1}^+ \tx{ and } \mathbf{O}_{+1}^- &&\tx{ where } &&(+1+1+1-1=2),
\end{align}
\begin{align}
    \mathbf{O}_{+2}^- \tx{ and } \mathbf{O}_{+2}^- &&\tx{ where } &&(+2+2-1-1=2),
\end{align}
\begin{align}
    \mathbf{O}_{+1}^- \tx{ and } \mathbf{O}_{+3}^- &&\tx{ where } &&(+1+3-1-1=2),
\end{align}
\begin{align}
    \mathbf{O}_{0}^- \tx{ and } \mathbf{O}_{+4}^- &&\tx{ where } &&(0+4-1-1=2),
\end{align}
\begin{align}
    \mathbf{O}_{0}^+ \tx{ and } \mathbf{O}_{-4}^+ &&\tx{ where } &&(0-4+1+1=-2).
\end{align}
\end{subequations}

Here the modes $\mathbf{O}_{0}^\pm=: \bf\Sigma^\pm$ correspond to circularly polarized modes without OAM ($m=0$). In particular, Eq.~\eqref{eq:first} reflects the process which individually conserves the SAM and OAM, and holds for any $m>0$. In Table~\ref{T2} in Appendix~\ref{sec:overlapintergalresults} we quantify the strength of this processes by calculating the corresponding overlap integral $I$ (Eq.~\eqref{eq:IOO}), and show that such processes dominate the cases in which the spin and orbital angular momenta are exchanged.

Besides the angular momentum conservation, the generation of allowed hybrid modes of signal and idler photons within the FWM process is defined by the usual phase-matching ($k_{1}+k_{2}-k_{s}-k_{i} \approx 0$) and frequency-matching ($\omega_{{1}}+\omega_{{2}}=\omega_{s}+\omega_i$) conditions.
Such conditions lead to producing FWM photons mostly with frequencies  $\omega_{s/i}=\omega_{{1/2}}$. However, in the presence of a helical grating, the phase-matching condition is modified by the grating's wave vector, which leads to a frequency shift of the generated FWM photons with respect to the frequencies of pumps. This will be demonstrated in the next subsections.


\subsection{Helical grating}\label{subsec:helical}



We now show how the presence of a helical, transmission grating~\cite{Erdogan} in the fiber can modify the conservation rules discussed above, allowing the generation higher-order OAM modes.
A sketch of a cylindrical step-index waveguide with a helical transmission grating and topological charges $m_g=1$ and $m_g=2$ is presented in Fig.~\ref{fig:grating}(c), where $\Lambda_{\mathrm{t}}$ is the grating period. The grating couples light of wavevector $k_\tx{in}$ with light of $k_\tx{out}= k_\tx{in} - 2 \pi/\Lambda_{\mathrm{t}}$. Current technologies~\cite{Fang:15, 21:Helical} allow to create a helical change of the refractive index in a waveguide core with a desired grating period and a non-zero topological charge. Such helical gratings can transform the fundamental waveguide mode $\mathbf{HE}_{1,1}^{(\eo)}$ into modes with non-zero orbital angular momentum with high efficiency.

We suppose that the waveguide core is characterized by a periodic change of its dielectric constant
\begin{equation}
\label{permitivity}
\Delta \varepsilon = \Delta \varepsilon_0  \cos (K_\tx{t} z - m_g \phi) \,\Theta (a-r),
\end{equation}
where $\Delta \varepsilon_0$ represents the modulation amplitude, $m_g$ is the topological charge of the grating, $a$ is the core radius and $K_\tx{t} = 2 \pi/\Lambda_\tx{t}$. The step-function $\Theta (a-r)$ indicates that the modulation of the permittivity is only applied to the core. Such a permittivity change leads to a continuous azimuthal variation of the refractive index along $z$.

Furthermore, since the helical gratings do not exhibit radial variation, we expect them not to  couple  orthogonal modes with different radial numbers $n$.


\subsubsection{Grating-controlled coupling between OAM modes}\label{sec:3B1}

For the forward-propagating signal photon, the coupling constant between the two coupled OAM modes $\mathbf{O}_{+m_s}^{\sigma_s}$ and  $\mathbf{O}^{\sigma'_s}_{+m'_s}$ is given by the transverse overlap integral:
\begin{align}
    \label{eq:kappa}
  &\kappa_s^{+m_s,\sigma_s; +m_s',\sigma_s'}(\omega_s)=
     \frac{\Delta\varepsilon_0\,\omega_s^2}{4k_s^{m_s}c^2} \\ \nn
     &\ \ \ \times
     \iint \tx{d}^2\mathbf{r}\,
     \qty[ \modeO{m_s}{\sigma_s}(r,\phi)]^{\!\,*} \!\cdot
     \modeO{m'_s}{\sigma'_s}(r,\phi)
    \; \Theta(a-r) \, e^{-i m_g \phi},
\end{align}
where $k_s^{m_s}$ is the wavevector of the $\mathbf{O}^{\sigma_s}_{m_s}$ mode, and $c$ is the speed of light. With the normalization introduced in Eq.~\eqref{eq:normalization}, the coupling coefficients are symmetric, i.e. $\kappa_s^{+m_s,\sigma_s; +m_s',\sigma_s'}=\kappa_s^{+m_s',\sigma_s'; +m_s,\sigma_s}$.

For the backward-propagating idler photon, the coupling constant $\kappa_i^{-m_i,\sigma_i; -m_i',\sigma_i'}(\omega_i)$ is determined by the same equation under substitutions $\modeO{+m_s}{\sigma_s} \rightarrow \modeO{-m_i}{\sigma_i}$, $\modeO{+m'_s}{\sigma'_s} \rightarrow \modeO{-m'_i}{\sigma'_i}$, and $k_s^{m_s}\rightarrow k_i^{-m_i}$.
Substituting the expressions of the transverse profiles of OAM modes given by Eqs.~\eqref{eq:Oplusdef} and~\eqref{eq:Ominusdef} into  Eq.~\eqref{eq:kappa}, and evaluating the azimuthal integrals ($\int \tx{d}\phi[\ldots]$), one can obtain the following selection rules for the topological charges:
\begin{subequations}
\begin{align}
    m'_s&=m_s+m_g && \tx{ if } \sigma_s=\sigma'_s,\label{eq:OAM_rule}\\
    m'_s&=m_s+m_g-2 && \tx{ if } \sigma_s=-\ \land \ \sigma'_s=+,\\
    m'_s&=m_s+m_g+2 && \tx{ if } \sigma_s=+\ \land \ \sigma'_s=-.
\end{align}
\end{subequations}

\begin{figure*}[hbtp]
  \includegraphics[width=.9\textwidth]{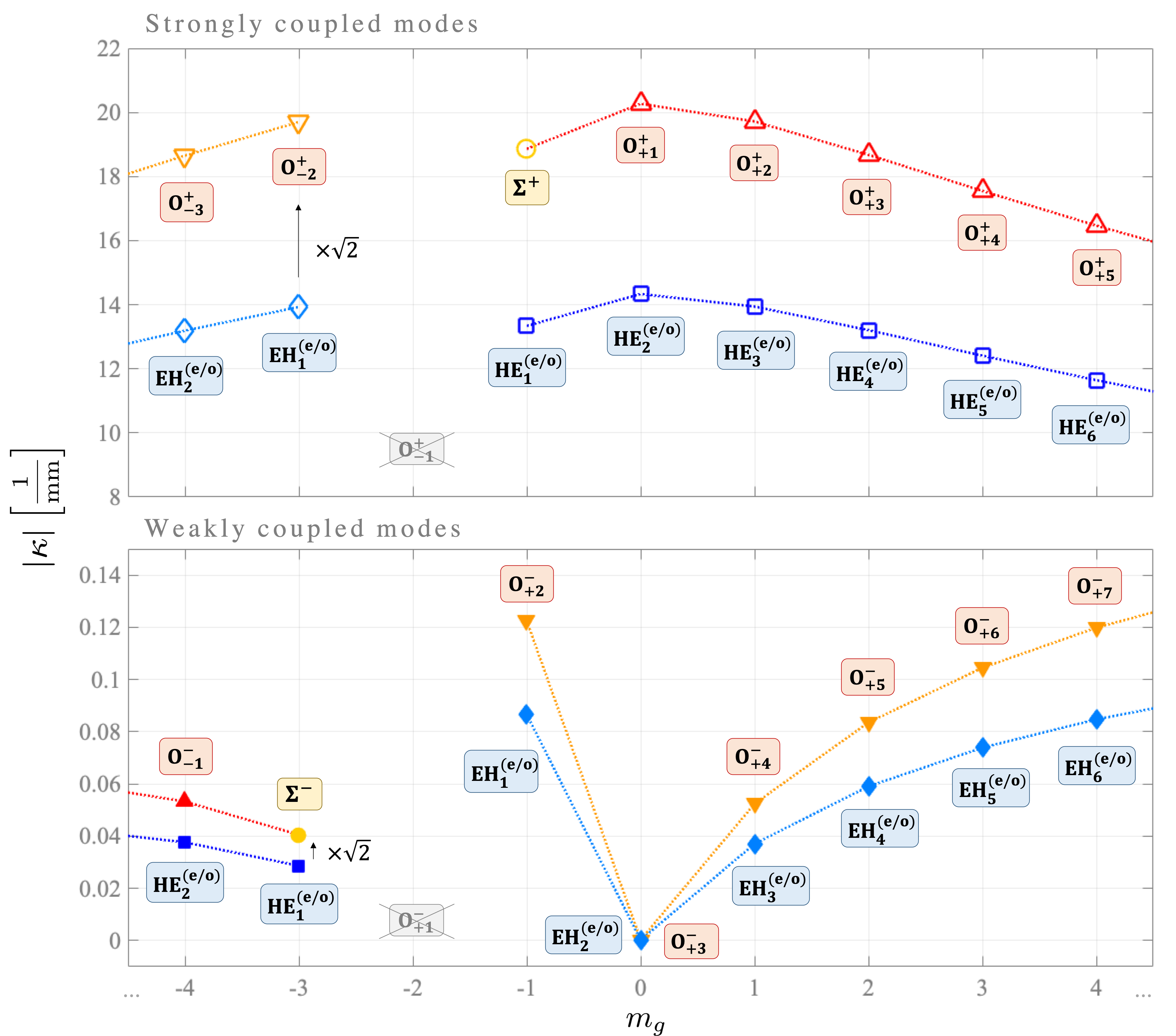}
        \caption{The absolute value of a coupling constant $|\kappa|$ between the input OAM mode $\mathbf{O}_{+1}^{\mathbf{+}}$ and all possible output hybrid and OAM modes calculated for a grating with $|m_g|$ spirals. The sign of $m_g$ determines the spirals' rotation direction ($m_g>0$ corresponds to a clockwise rotation direction and $m_g<0$ is associated with a counterclockwise rotation).
        As the $\textbf{O}^+_{-1}$ and $\textbf{O}^-_{+1}$ modes are formed by unstable TE and TM modes, they are not further considered and are marked by gray, crossed out boxes on the bottom of the corresponding plot at $m_g=-2$. 
        }
    \label{fig:cc}
\end{figure*}

To illustrate these OAM conservation rules, the coupling constants between the input mode $\mathbf{O}_{+1}^{+}$ and the phase-matched OAM  and hybrid modes are shown in Fig.~\ref{fig:cc}.
In calculations, a low-contrast silica optical waveguide  is considered with refractive indices of core and cladding, $n_\tx{co} = 1.45$ and $n_\tx{cl} = 1.44$, respectively. The perturbation strength of the grating is $ \Delta n_0 = 0.01$, the core radius $a = 20\,$\textmu$ \mathrm{m}$, the pump wavelength $\lambda = 1.55\,$\textmu$\mathrm{m}$, $\Delta \varepsilon_0 \approx 2 n_{\mathrm{co}} \Delta n_0 = 2.9 \times 10^{-2}$, and the parameter $V={2 \pi a} \sqrt{\varepsilon_\mathrm{co}-\varepsilon_\mathrm{cl}}/{\lambda} \approx 13.8$, where $\varepsilon_\tx{co}$ and $\varepsilon_\tx{cl}$ are the dielectric constants in the core and cladding, respectively.

Figure~\ref{fig:cc} indicates how one can properly choose the grating order $m_g$ to couple an input OAM mode to the desired output OAM mode.
In the top panel, the output modes carry the same spin, $\sigma=+$, as the input mode $\mathbf{O}_{+1}^{+}$\footnote{Note that OAM modes with spin different to the input mode are also coupled due to the spin-orbit-coupling effect, but are not further discussed nor listed, since the coupling is very weak, with $\kappa$ of order $0.1$/cm.} their orbital angular momentum number $m'$ is modified by the helical charge of the grating as $m'=m+m_g$, where here $m=1$ is the input mode's orbital number (see selection rule~\eqref{eq:OAM_rule}).

In the bottom panel we consider the coupling to OAM modes with opposite polarization, $\sigma=-$, in comparison to the input mode $\mathbf{O}_{+1}^{+}$, and find that the resulting coupling constants are much smaller by a factor of $10^{-2}$.



In the following discussions we will employ a convention in which un-primed orbital numbers ($m_s$ and $m_i$)  describe photons locally generated by the FWM process, while  primed orbital numbers ($m_s'$ and $m_i'$) denote the OAM of photons generated by the grating-mediated coupling to the higher OAM modes.

\subsection{Asymptotic-in and -out formalism}\label{subsec:asymptotic}

As photons are being generated through the FWM and propagate along the fiber, they are being continuously and selectively converted between OAM modes by the helical grating. To describe the spatial distribution of the signal and idler photons, we embrace the  \textit{asymptotic-input and output modes} formalism developed by Liscidini \textit{et al.}~\cite{asymptotic}. Within this approach, the helical gratings define the grating modes of the signal (idler) photons, made up of oscillating superpositions of the modes with $m$ and $m'$ OAM charges. To complete the definition of these grating modes, we impose boundary conditions which state that the \textit{photons emerge from the waveguide with either a dominating $\mathbf{O}^{\sigma_s}_{m_s}$ $\qty(\mathbf{O}^{\sigma_i}_{-m_i})$ or $\mathbf{O}^{\sigma'_s}_{m'_s}$ $\qty(\mathbf{O}^{\sigma'_i}_{-m'_i})$ mode profile}. The polarizations of these coupled modes are set to be the same throughout this subsection (i.e. $\sigma_s=\sigma'_s$ and $\sigma_i=\sigma'_i$), and the primed orbital numbers $m_s'$ and $m_i'$ will be uniquely determined by the grating structure according to the selection rule~\eqref{eq:OAM_rule}.

\subsubsection{Asymptotic-out modes of the signal photon}

We first consider the signal photon propagating in the $+z$ direction, and emerging from the right end of the waveguide ($z=L$) in mode $(m_s,\sigma_s)$, after it oscillates between the $(m_s,\sigma_s)$ and $(m_s',\sigma_s)$ modes due to the grating.
The corresponding electric field operator $\hat{\bf E}_s^{(m_s,\sigma_s)}$ of that asymptotic-out mode is
\begin{widetext}
\begin{align}
   \hat{\bf E}_s^{(m_s,\sigma_s)}(\mathbf{r},t)
    &=
      \  \int   \textup{d}\omega_s \  \mathcal{N}_s(\omega_s)
           \Bigg[
           a_{m_s\rightarrow m_s}(z) \; \modeO{m_s}{\sigma_s}(r,\phi) \;e^{ik_s^{m_s}(\omega_s)z} \nn\\
           &\qquad\qquad\qquad\qquad + a_{m_s'\rightarrow m_s}(z) \; \modeO{m_s'}{\sigma_s}(r,\phi) \;e^{ik_s^{m_s'}(\omega_s)z} \Bigg] \
           \hat{b}_s^{\tx{O}_{m_s}^{\sigma_s}}(\omega_s)\;
           e^{-i\omega_s t}\,
           +\textup{h.c.},
           \nn\\
           & =   \  \int   \textup{d}\omega_s \  \mathcal{N}_s(\omega_s) \Bigg[ \sum_{y_s=m_s, m'_s} a_{y_s\rightarrow m_s}(z) \; \modeO{y_s}{\sigma_s}(r,\phi) \;e^{ik_s^{y_s}(\omega_s)z}  \Bigg] \ \hat{b}_s^{\tx{O}_{m_s}^{\sigma_s}}(\omega_s)\;
           e^{-i\omega_s t}\,
           +\textup{h.c.},
\label{eq:Es}
\end{align}
\end{widetext}
where the operator $\qty(\hat{b}_s^{\mathrm{O}_{m_s}^{\sigma_s}})^{\!\dagger}$ describes the creation of the signal photon with topological charge $m_s\in\mathbb{N}_+$ at the output of the fiber. In the above formulation, we implicitly assume that all the functions in the integrand are parametrically dependent on $\omega_s$.

Considering the transverse modes normalized to the Pointing vector according to Eq.~\eqref{eq:normalization}, the field normalization factor is given by 

\begin{align}
    \mathcal{N}_{s}(\omega_s) = \sqrt{\frac{\hbar\,\omega_{s} }{\pi\, (\varepsilon_{co}+\varepsilon_{cl})} },
    \label{eq:norm}
\end{align}
where the expression from Ref.~\cite{Ebers_2022} was applied taking into account that in our case  $S_{\mathrm{eff}_s} = \frac{(\varepsilon_{co}+\varepsilon_{cl})n_{\tx{eff}_s}(\omega_{s})}{4 \varepsilon_0 c} $ with $n_{\tx{eff}_s}(\omega_{s})$ labels the frequency-dependent effective refractive index of mode.

In Eq.~\eqref{eq:Es}, the slowly varying amplitudes $a_{m_s\rightarrow m_s}(z)$ and $a_{m'_s\rightarrow m_s}(z)$ are defined as solutions to the grating-coupled mode equations:
\begin{widetext}
\begin{subequations}
\begin{align}
\label{coupled.modes.HE}
    i\frac{\text{d}}{\text{d}z} a_{m_s\rightarrow m_s}(z)
    +\frac{\omega_s-\omega_{t,s}}{v_g^{m_s}}\; a_{m_s\rightarrow m_s}(z)
    + \kappa_s \; a_{m'_s\rightarrow m_s}(z) \;e^{i\qty[ k_s^{m'_s}(\omega_s) -k_s^{m_s}(\omega_s)-K_s]z} =0,\\
    i\frac{\text{d}}{\text{d}z} a_{m'_s\rightarrow m_s}(z)
    +\frac{\omega_s-\omega_{t,s}}{v_g^{m'_s}}\; a_{m'_s\rightarrow m_s}(z)
    + \kappa_s^* \;a_{m_s\rightarrow m_s}(z) \;e^{-i\qty[k_s^{m'_s}(\omega_s) -k_s^{m_s}(\omega_s)-K_s]z} =0.
\end{align}\label{CME}
\end{subequations}
\end{widetext}
In the above, we simplify the notation using $\kappa_s$ as the appropriate coupling coefficient defined in Eq.~\eqref{eq:kappa}. We also account for both the frequency detuning between the transmission grating frequency $\omega_{t,s}$ and the signal photon frequency $\omega_s$, $\delta_s=\omega_{t,s}-\omega_s$, and the momentum detuning $2d_s =\delta_s\qty(1/v_{g}^{m'_s}-1/v_{g}^{m_s})+K_s+k_s^{m_s}(\omega_s)-k_s^{m'_s}(\omega_s)$, using the group velocities $v_{g}^{m_s/m'_s}$ at the signal frequency $\omega_s$.

To identify the asymptotic-out modes associated with the generation of signal photon in the  $\mathbf{O}^{\sigma_s}_{m_s}$ mode, we add the boundary conditions $a_{m_s\rightarrow m_s} (z=L) = 1$ and $a_{m'_s\rightarrow m_s} (z=L)=0$, and find the general solutions for the mode amplitudes:
\begin{subequations}
    \label{eq:as1}
    \begin{align}
    a_{m_s\rightarrow m_s}(z)
    &= e^{i(L-z)\qty(d_s+\delta_s/v_g^{m_s})}\nn\\  &\ \ \ \times \left\{\cos[\gamma_s(L-z)]-i\frac{d_s}{\gamma_s}\sin\qty[\gamma_s(L-z)]\right\},\\
    a_{m'_s\rightarrow m_s}(z)
    &= -i\,\frac{\kappa_s^*}{\gamma_s}\,e^{i \qty[(L+z) d_s-\delta_s\qty(z/v_g^{m'_s}-L/v_g^{m_s})]} \nn\\  &\ \ \ \times\sin[\gamma_s(L-z)],
\end{align}
\end{subequations}
\!\!where 
\begin{equation}\label{eq:param}
    \gamma_s=\sqrt{d_s^2+|\kappa_s|^2}.
\end{equation}
We should note that the exact, constant phase of the envelope functions is irrelevant for calculating the joint spectral intensity (JSI), since it relies on the arbitrary choice of time $t$, and we do not consider any interferences between competing nonlinear processes.

{In the above model, we assumed that the signal photon would oscillate between modes with topological charge $m_s$ and $m_s'$ (where $m_s'=m_s+m_g$), and emerge in the former mode. To complement this description, we define operator $\hat{\bf E}_s^{(m_s',\sigma_s)}$ similarly to Eq.~\eqref{eq:Es}, but which describes the generation of the output photon in the $\mathbf{O}^{\sigma_s}_{m'_s}$ mode at the end of the waveguide:
\begin{widetext}
\begin{align}
   \hat{\bf E}_s^{(m_s',\sigma_s)}(\mathbf{r},t)
    &=
      \   \int   \textup{d}\omega_s \ \mathcal{N}_s(\omega_s)
           \Bigg[
           a_{m_s\rightarrow m_s'}(z) \; \modeO{m_s}{\sigma_s}(r,\phi) \;e^{ik_s^{m_s}(\omega_s)z} \nn\\
           &\qquad\qquad\qquad\qquad + a_{m_s'\rightarrow m_s'}(z) \; \modeO{m_s'}{\sigma_s}(r,\phi) \;e^{ik_s^{m_s'}(\omega_s)z} \Bigg] \
           \hat{b}_s^{\tx{O}_{m_s'}^{\sigma_s}}(\omega_s)\;
           e^{-i\omega_s t}\,
           +\textup{h.c.}
           \nn\\
           & =   \  \int   \textup{d}\omega_s \  \mathcal{N}_s(\omega_s) \Bigg[ \sum_{y_s=m_s, m'_s} a_{y_s\rightarrow m'_s}(z) \; \modeO{y_s}{\sigma_s}(r,\phi) \;e^{ik_s^{y_s}(\omega_s)z}  \Bigg] \ \hat{b}_s^{\tx{O}_{m'_s}^{\sigma_s}}(\omega_s)\;
           e^{-i\omega_s t}\,
           +\textup{h.c.}.
\label{eq:Esp}
\end{align}
\end{widetext}
We also modify the boundary conditions: $a_{m_s\rightarrow m_s'}(z=L) = 0$ and $a_{m'_s\rightarrow m'_s}(z=L) = 1$ in the similarly modified grating-coupled mode equations.}
The explicit expressions for the amplitudes $a_{m_s\rightarrow m_s'}(z)$ and $a_{m'_s\rightarrow m_s'}(z)$  of the signal photon are shown in Appendix~\ref{sec:input_output}.


We acknowledge that this formulation introduces some ambiguity. For example, given the grating number $m_g=2$, the operator $\hat{\bf E}_s^{(3,+)}$ can conceivably describe a signal photon with, either, $m_s=3$ generated via a coupling with $m_s'=5$ (see Eq.~\eqref{eq:Es}), or with $m_s'=3$ generated via the coupling with $m_s=1$ (see Eq.~\eqref{eq:Esp}). This is resolved by assuming that the FWM process will predominantly generate photons with low topological charges $m$, and therefore the latter process will be dominant.

\subsubsection{Asymptotic-out modes of the idler photon}

The left-propagating asymptotic-out idler modes are given by an analogous formulation for the electric field operators:
\begin{widetext}
\begin{align}
   \hat{\bf E}^{(-m_i,\sigma_i)}_i(\mathbf{r},t)
    &=
      \int   \textup{d}\omega_i \ \mathcal{N}_i(\omega_i)
           \Bigg[
           a_{-m_i\leftarrow -m_i}(z) \; \modeO{-m_i}{\sigma_i}(r,\phi) \;e^{-ik_i^{-m_i}(\omega_s)z} \nn\\
           &\qquad\qquad\qquad\qquad + a_{-m_i\leftarrow -m_i'}(z) \; \modeO{-m_i'}{\sigma_i}(r,\phi) \;e^{-ik_i^{-m_i'}(\omega_i)z} \Bigg] \
           \hat{b}_i^{\tx{O}_{-m_i}^{\sigma_i}}(\omega_i)\;
           e^{-i\omega_i t}\,
           +\textup{h.c.}
           \nn\\
           & =   \  \int   \textup{d}\omega_i \  \mathcal{N}_i(\omega_i) \Bigg[ \sum_{y_i=m_i, m'_i} a_{-m_i\leftarrow -y_i}(z) \; \modeO{-y_i}{\sigma_i}(r,\phi) \;e^{-ik_i^{y_i}(\omega_s)z}  \Bigg] \hat{b}_i^{\tx{O}_{-m_i}^{\sigma_i}}(\omega_i)\;
           e^{-i\omega_i t}\,
           +\textup{h.c.},
\label{eq:Ei}
\end{align}
\begin{align}
   \hat{\bf E}^{(-m_i',\sigma_i)}_i(\mathbf{r},t)
    &=
      \int   \textup{d}\omega_i \ \mathcal{N}_i(\omega_i)
           \Bigg[
           a_{-m_i'\leftarrow -m_i}(z) \; \modeO{-m_i}{\sigma_i}(r,\phi) \;e^{-ik_i^{-m_i}(\omega_s)z} \nn\\
           &\qquad\qquad\qquad\qquad + a_{-m_i'\leftarrow -m_i'}(z) \; \modeO{-m_i'}{\sigma_i}(r,\phi) \;e^{-ik_i^{-m_i'}(\omega_i)z} \Bigg] \
           \hat{b}_i^{\tx{O}_{-m_i'}^{\sigma_i}}(\omega_i)\;
           e^{-i\omega_i t}\,
           +\textup{h.c.}
           \nn\\
           & =   \  \int   \textup{d}\omega_i \  \mathcal{N}_i(\omega_i) \Bigg[ \sum_{y_i=m_i, m'_i} a_{-m'_i\leftarrow -y_i}(z) \; \modeO{-y_i}{\sigma_i}(r,\phi) \;e^{-ik_i^{y_i}(\omega_s)z}  \Bigg] \ \hat{b}_i^{\tx{O}_{-m'_i}^{\sigma_i}}(\omega_i)\;
           e^{-i\omega_i t}\,
           +\textup{h.c.},
\label{eq:Eip}
\end{align}
\end{widetext}
with the corresponding envelope functions given in Appendix~\ref{sec:input_output}. The orbital numbers $-m_i$ and $-m'_i$ are by construction negative so that they possess an opposite sign in comparison to the orbital numbers $m_s$ and $m'_s$ of the signal mode; the total angular momentum conservation rule~\eqref{eq:mcondition} is thus satisfied, where no spin-orbit coupling effect during the FWM process is considered.




\subsubsection{Asymptotic-input modes of the pumps}

The electric fields associated with the two pump beams are described as classical, non-depleted continuous-wave (CW) pump fields are given as
\begin{subequations}
\begin{align}
	\textbf{E}_{1} &=
	    E_0\;
        \bcalE_{\!p_1}(r,\phi) \;
        e^{i(k_1 z -\omega_1 t)}
        +   \textup{c.c.},
        \\
	\textbf{E}_{2}
    	&=
	    E_0\;
        \bcalE_{\!p_2}(r,\phi)\;
        e^{i(-k_2 z -\omega_2 t)}
        +   \textup{c.c.}.
\end{align}\label{eq:Ep2}
\end{subequations}
\!\!\! In the waveguide the pumps have identical field amplitudes $E_0$ and exhibit transverse mode profiles corresponding to the fundamental waveguide mode $\textbf{HE}_{1}^{(\ee)}$. The frequencies $\omega_{1,2}$ are assumed to be far from resonance with the transmission gratings' frequencies.
The first pump is forwards propagating with $z$-projection of the wavenumber $k_1>0$, the second is backwards propagating with $k_2<0$.



\section{OAM pair generation}\label{sec:generation}


{The electric field operator describing the signal (and idler) photons can be thus expressed as a superposition of the expressions given in Eq.~\eqref{eq:Es} and Eq.~\eqref{eq:Esp} (Eq.~\eqref{eq:Ei} and Eq.~\eqref{eq:Eip}). Note that each of these expressions connects only two orbital numbers: $m_s$ and $m'_s$ or $m_i$ and $m'_i$, respectively. Therefore, to account for all possible OAM modes that can be generated in such a process, we sum the fields over $m_s$ and $m_i$ (see note above about the solution to the ambiguity of notation):
\begin{subequations}
\begin{align}
   \hat{\bf E}_s^{(\sigma_s)} &= \sum_{m_s>0} \ \  \sum_{x_s=m_s,m_s'>0} \hat{\bf E}_s^{(x_s,\sigma_s)},
   \\
   ~\hat{\bf E}_i^{(\sigma_i)} &= \sum_{m_i>0} \ \ \sum_{x_i=m_i,m_i'>0} \hat{\bf E}_i^{(-x_i,\sigma_i)},
\end{align}
\label{eq:EsEi.total}\end{subequations}
where the values $m'_s$ and $m'_i$ are again assumed to be determined by the grating selection rules; see Eq.~\eqref{eq:OAM_rule}.
}

We can thus rewrite the nonlinear Hamiltonian given in Eq.~\eqref{eq:H} as:
\begin{widetext}
\begin{align}\label{Hamiltonian.02}
\hat H_{_\mathrm{NL}}= \;&
   \varepsilon_0 \sum_{\sigma_s,\sigma_i=\pm} \iint \mathrm{d} \omega_s \mathrm{d} \omega_i \int_V \mathrm{d}^3 \textbf{r}\  \chi^{(3)} \,
    \textbf{E}_{1}^{(+)} \,
    \textbf{E}_{2}^{(+)} \,
    \left(\hat{\textbf{E}}_{s}^{(\sigma_s)}\right)^{(-)} \,
    \left(\hat{\textbf{E}}_{i}^{(\sigma_i)}\right)^{(-)} \,
    +{\textup{h.c.}} \nn\\
 = \;&
 \hbar \sum_{\sigma_s,\sigma_i=\pm}\;
 \sum_{m_s, m_i, >0}\;
\sum_{x_s=m_s, m'_s >0}\;
\sum_{x_i=m_i, m'_i >0}\;
\sum_{y_s=m_s, m'_s >0}\;
\sum_{y_i=m_i, m'_i >0}\;
 \iint \mathrm{d} \omega_s \mathrm{d} \omega_i \ \tilde{\Gamma}(\omega_s,\omega_i) \int_0^L\mathrm{d}z  \ e^{-i\,\delta\omega\,t} \nn \\
& \times I^{y_s,\sigma_s;-y_i,\sigma_i} \
         \qty[a_{y_s\rightarrow x_s}(z)]^* \; \qty[a_{-x_i \leftarrow -y_i}(z)]^*\
         e^{i\qty[k_1-k_2- k_s^{y_s}+ k_i^{-y_i}]z}\
         \qty[\hat b_s^{\tx{O}_{x_s}^{\sigma_s}}]^\dagger\qty[\hat b_i^{\tx{O}_{-x_i}^{\sigma_i}}]^\dagger +\tx{h.c.},
\end{align}
with $\tilde{\Gamma}(\omega_s,\omega_i)= \varepsilon_0 \, \mathcal{N}_s(\omega_s) \, \mathcal{N}_i(\omega_i) \, E_0^2 /  \hbar$, where $\mathcal{N}_{s}(\omega_s)$ is defined by Eq.~\eqref{eq:norm} and $\mathcal{N}_{i}(\omega_i)$ by the same equation after substitution of the frequency of the signal photon with those of the idler photon: $\omega_s\rightarrow\omega_i$. $\delta \omega = \omega_1 + \omega_2 - \omega_s -\omega_i$ defines the frequency mismatch of the four modes, and the overlap integral $I^{n_s,\sigma_s;n_i,\sigma_i}$ is given by Eq.~\eqref{eq:IOO}.

We can now single out several terms in the above expression which describe the generation of photon pairs from the FWM process involving low-OAM modes, namely, ($m_s, m_i$), as these terms exhibit the greatest contribution due to the largest overlap integral:
\begin{align}
\hat H_{_\mathrm{NL}}\approx \;&
\hbar \sum_{\sigma_s,\sigma_i=\pm}\;
\sum_{m_s,m_i>0}\;
 \iint \mathrm{d} \omega_s \mathrm{d} \omega_i \ \tilde{\Gamma}(\omega_s,\omega_i) \int_0^L\mathrm{d}z  \ e^{-i \delta\omega t} I^{m_s,\sigma_s;-m_i,\sigma_i} \bigg\{ \nn \\
&  \
         \qty[a_{m_s\rightarrow m_s}(z)]^* \; \qty[a_{-m_i \leftarrow -m_i}(z)]^*\
         e^{i\qty[k_1-k_2- k_s^{m_s}+ k_i^{-m_i}]z}\
         \qty[\hat b_s^{\tx{O}_{m_s}^{\sigma_s}}]^\dagger\qty[\hat b_i^{\tx{O}_{-m_i}^{\sigma_i}}]^\dagger \nn \\
& + \
         \qty[a_{m_s\rightarrow m_s}(z)]^* \; \qty[a_{-m_i' \leftarrow -m_i}(z)]^*\
         e^{i\qty[k_1-k_2- k_s^{m_s}+ k_i^{-m_i}]z}\
         \qty[\hat b_s^{\tx{O}_{m_s}^{\sigma_s}}]^\dagger\qty[\hat b_i^{\tx{O}_{-m_i'}^{\sigma_i}}]^\dagger \nn \\
& + \
         \qty[a_{m_s\rightarrow m_s'}(z)]^* \; \qty[a_{-m_i \leftarrow -m_i}(z)]^*\
         e^{i\qty[k_1-k_2- k_s^{m_s}+ k_i^{-m_i}]z}\
         \qty[\hat b_s^{\tx{O}_{m_s'}^{\sigma_s}}]^\dagger\qty[\hat b_i^{\tx{O}_{-m_i}^{\sigma_i}}]^\dagger \nn \\
& + \
         \qty[a_{m_s\rightarrow m_s'}(z)]^* \; \qty[a_{-m_i' \leftarrow -m_i}(z)]^*\
         e^{i\qty[k_1-k_2- k_s^{m_s}+ k_i^{-m_i}]z}\
         \qty[\hat b_s^{\tx{O}_{m_s'}^{\sigma_s}}]^\dagger\qty[\hat b_i^{\tx{O}_{-m_i'}^{\sigma_i}}]^\dagger  \bigg\}+\tx{h.c.}.\label{Hamiltonian03}
\end{align}
\end{widetext}

This form recognizes explicitly that the four dominant pairs of photons will all be generated by the local FWM processes involving the strong pumps, and low-OAM modes $m_s$ and $m_i$.

For reference, we note that in the absence of gratings, the expressions for the electric field operators of the signal photon can be represented as in Eqs.~\eqref{eq:Es} and Eq.~\eqref{eq:Esp} but with amplitudes $a_{m_s\rightarrow m_s}(z)=a_{m_s'\rightarrow m_s'}(z)=1$ and $a_{m_s'\rightarrow m_s}(z)=a_{m_s\rightarrow m_s'}(z)=0$, showing the lack of oscillation between different OAM modes. A similar formulation holds for the idler photons, such that
the total Hamiltonian can be simplified to:
\begin{widetext}
\begin{align}\label{H0}
\hat H_{_{\mathrm{NL}_0}}=\;&
\hbar \sum_{\sigma_s,\sigma_i=\pm}\;
\sum_{m_s,m_i>0}\;
 \iint \mathrm{d} \omega_s \mathrm{d} \omega_i \ \tilde{\Gamma}(\omega_s,\omega_i) \int_0^L\mathrm{d}z  \ e^{-i\,\delta\omega\,t} \nn \\
&\qquad\qquad \times I^{m_s,\sigma_s;-m_i,\sigma_i} \
         e^{i\qty[k_1-k_2- k_s^{m_s}+ k_i^{-m_i}]z}\
         \qty[\hat b_s^{\tx{O}_{m_s}^{\sigma_s}}]^\dagger\qty[\hat b_i^{\tx{O}_{-m_i}^{\sigma_i}}]^\dagger +\tx{h.c.}.
\end{align}
\end{widetext}


\subsection{Calculating the Joint Spectral Amplitudes}\label{subsec:jsa}

Spectra of light generated through these nonlinear processes are characterized by the Joint Spectra Amplitudes (JSAs) $\Phi^{x_s,\sigma_s;-x_i,\sigma_i}(\omega_s,\omega_i)$ and
Intensities (JSIs) $\qty|\Phi^{x_s,\sigma_s;-x_i,\sigma_i}(\omega_s,\omega_i)|^2$. Here, $x_{s}\in\{+m_{s},+m'_{s}\}$ and $x_i\in\{-m_i,-m'_i\}$ describe the topological charge of the emitted signal and idler photons, respectively.

To determine the JSAs, the first-order perturbation in $\hat H_{_\mathrm{NL}}$ is used to find the state generated by the FWM process, where we relate the amplitudes of each generated OAM mode pair with the specific JSA:
\begin{widetext}
\begin{align}
    \ket{\psi_{\tx{gen}}} &=
         -i \int \mathrm{d}\omega_s\;\Gamma(\omega_s,\bar{\omega}_i)
        \nn \\
        &\qquad \times \sum_{\sigma_s,\sigma_i=\pm}\;
\sum_{m_s,m_i>0}\;
         \sum_{x_s=m_s, m'_s, >0}\;
\sum_{x_i=m_i, m'_i, >0} \Phi^{x_s,\sigma_s;-x_i,\sigma_i}(\omega_s,\bar{\omega}_i)
            \qty[\hat b_{s}^{\tx{O}_{x_s}^{\sigma_s}}(\omega_s)]^{\!\dagger} \;
            \qty[\hat b_{i}^{\tx{O}_{-x_i}^{\sigma_i}}(\bar{\omega}_i)]^{\!\dagger}
            \, \vert 0 \rangle_s \vert 0 \rangle_i,
        \label{JSA_total}
\end{align}
with the two-photon amplitudes
\begin{align}\label{jsa.oam2}
\Phi^{x_s,\sigma_s;-x_i,\sigma_i}=
  \sum_{y_s=m_s, m'_s >0}\;
\sum_{y_i=m_i, m'_i >0}\;
      I^{y_s,\sigma_s;-y_i,\sigma_i}\,
     \int_0^L \mathrm{d}z \ e^{i\left(k_1-k_2- k_s^{y_s}
      + k_i^{-y_i} \right)z}\left[a_{y_s\rightarrow x_s} (z) \right]^*
      \left[a_{-x_i \leftarrow -y_i} (z)\right]^*.
\end{align}
\end{widetext}

Here, we have already supposed a CW pumping scheme, such that the integral over time results in a perfect matching of the frequencies of the signal and idler photons: $\bar{\omega}_i = \omega_1+\omega_2 - \omega_s$.
 
Focusing on the terms included in the simplified Hamiltonian given in Eq.~\eqref{Hamiltonian03}, we find
\begin{widetext}
\begin{align}
    \ket{\psi_{\tx{gen}}} \approx 
         \int&\mathrm{d}\omega_s\;\Gamma(\omega_s,\bar{\omega}_i) \;
         \sum_{\sigma_s,\sigma_i=\pm} \ \
         \sum_{m_s,m_i>0}\;
        \Bigg\{ \nn\\
    		&\Phi^{m_s,\sigma_s;-m_i,\sigma_i}(\omega_s,\bar{\omega}_i)\;
            \qty[\hat b_{s}^{\tx{O}_{m_s}^{\sigma_s}}(\omega_s)]^{\!\dagger} \;
            \qty[\hat b_{i}^{\tx{O}_{-m_i}^{\sigma_i}}(\bar{\omega}_i)]^{\!\dagger}\nn\\
    		&+
            \Phi^{m_s,\sigma_s;-m_i',\sigma_i}(\omega_s,\bar{\omega}_i)\;
            \qty[\hat b_{s}^{\tx{O}_{m_s}^{\sigma_s}}(\omega_s)]^{\!\dagger} \;
            \qty[\hat b_{i}^{\tx{O}_{-m_i'}^{\sigma_i}}(\bar{\omega}_i)]^{\!\dagger}
            \nn \\
    		&+
            \Phi^{m_s',\sigma_s;-m_i,\sigma_i}(\omega_s,\bar{\omega}_i)\;
            \qty[\hat b_{s}^{\tx{O}_{m_s'}^{\sigma_s}}(\omega_s)]^{\!\dagger} \;
            \qty[\hat b_{i}^{\tx{O}_{-m_i}^{\sigma_i}}(\bar{\omega}_i)]^{\!\dagger}
            \nn\\
            &+
    	\Phi^{m_s',\sigma_s;-m_i',\sigma_i}(\omega_s,\bar{\omega}_i)\;
            \qty[\hat b_{s}^{\tx{O}_{m_s'}^{\sigma_s}}(\omega_s)]^{\!\dagger}\;
            \qty[\hat b_{i}^{\tx{O}_{-m_i'}^{\sigma_i}}(\bar{\omega}_i)]^{\!\dagger}\;
        \Bigg\}
            \, \vert 0 \rangle_s \vert 0 \rangle_i.
        \label{JSA_totalb}
\end{align}
\end{widetext}

Similarly, considering the nonlinear Hamiltonian $\hat H_{_{\tx{NL}_0}}$ in Eq.~\eqref{H0} describing system without the gratings, we can find the output state
\begin{align}\label{eq:state_0}
    \ket{\psi_{\mathrm{gen}_0}}
    &= 
        \int
        \mathrm{d}\omega_s \;\Gamma(\omega_s,\bar{\omega}_i)
        \nn\\&
        \ \ \ \times
        {\sum_{\sigma_s,\sigma_i=\pm}}\
        {\sum_{m_s,m_i>0}}\Bigg\{ \Phi^{m_s,\sigma_s;-m_i,\sigma_i}_0 (\omega_{s},\bar{\omega}_{i})  \nn\\
            &\ \ \ \times \qty[\hat b_{s}^{\tx{O}_{m_s}^{\sigma_s}}(\omega_s)]^{\!\dagger} \,
            \qty[\hat b_{i}^{\tx{O}_{-m_i}^{\sigma_i}}(\bar{\omega}_i)]^{\!\dagger}  \Bigg\}
             \ket{0}_s  \ket{0}_i
,
\end{align}
with the JSA of the process
\begin{align}
    &\Phi^{m_s,\sigma_s;-m_i,\sigma_i}_0 (\omega_{s},\bar{\omega}_{i})  \nn\\
    &\ =
     I^{m_s,\sigma_s;-m_i,\sigma_i}
             \int_0^L \mathrm{d}z \  e^{i\left[k_1-k_2- k_s^{m_s}(\omega_s)
              + k_i^{-m_i}(\bar{\omega}_i) \right]z}.
    \label{JSA_0}
\end{align}

In the following, we will for simplicity omit the superscripts of $\Phi$ indicating the SAM $\sigma_{s,i}$, as the handedness of $\sigma_{s,i}$ is crucial for the phase-matching condition in Eq.~\eqref{eq:mcondition}, but does not change the JSAs once the condition is fulfilled. Furthermore, any spin-orbit coupling effects during the FWM process are neglected due its associated low strength.

\subsection{Numerical results}\label{subsec:numerical.results}

\begin{figure}[ht]
  \includegraphics[width=\columnwidth]{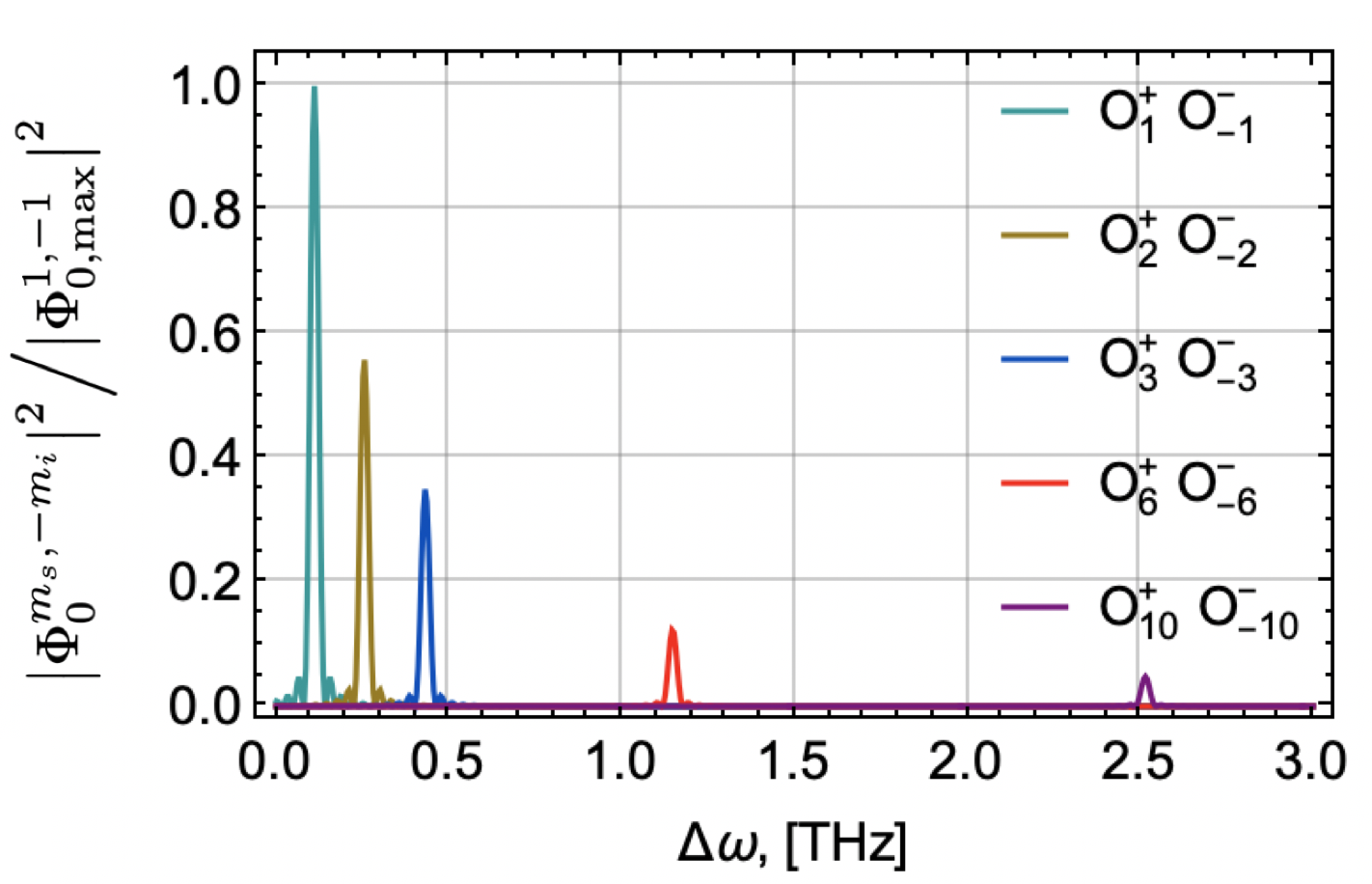}
        \caption{Normalized joint spectral intensities (JSIs) $\qty|\Phi_0^{m_s,-m_i}|^2\Big/\qty|\Phi_\tx{0,max}^{1,-1}|^2$ describing the generation of pairs of photons in OAM modes $\mathbf O^+_{m_s}$ and $\mathbf O^-_{-m_i}$ during the FWM process in a waveguide without a grating. The normalization is preformed with respect to the maximum of JSI $\qty|\Phi^{1,-1}_\tx{0,max}|^2 $ describing the generation of a pair of photons in modes $\mathbf O^+_{1}$ and $\mathbf O^-_{-1}$. Due to the decrease in the overlap integrals, the probability of generating non-zero OAM modes decreases with an increase in the topological charge $m_{s,i}$ of the generated modes.
        $\Delta \omega = \omega_s - \omega_{1} = \omega_{2} - \bar{\omega}_i$ is the frequency detuning of the modes, see Fig.~\ref{fig:grating}(a). The pump modes are both in the fundamental mode $\mathbf{HE}_1^{(\ee)}$ and exhibit the wavelengths $\lambda_1 = 1.5 \, \upmu$m and $\lambda_2 = 0.5 \, \upmu$m, respectively, the  waveguide length is 2\,cm.  These parameters are fixed for all other plots and tables.}
    \label{fig:JSI_without_grating}
\end{figure}

We start our analysis by considering the FWM process in a waveguide without the grating. Fig.~\ref{fig:JSI_without_grating} shows the corresponding normalized JSIs $|\Phi^{m_s,\sigma_s;-m_i,\sigma_i}_0|^2$, see Eq.~\eqref{JSA_0}, which describe the generation of pairs of photons in the OAM modes $\mathbf{O}^+_{+m_s}$ and $\mathbf{O}^-_{-m_i}$ for various $m_s$ and $m_i$ assuming both pump photons exhibit the fundamental $\mathbf{HE}_1^{(\ee)}$ mode profile. In this figure, the JSIs are plotted as a function of detuning $\Delta\omega=\omega_s-\omega_1 = \omega_2-\bar{\omega}_i$, and normalized to the maximum value of JSI $\qty|\Phi_\tx{0,max}^{1,-1}|^2$, associated with the generation of a pair of photons in the modes $\mathbf O^+_{+1}$ and $\mathbf O^-_{-1}$. One can clearly observe that the generation efficiency of the OAM modes decreases with increasing absolute orbital number $m_s=m_i$ --- the generation of high-order OAM modes in conventional waveguides is thus problematic.

As discussed earlier, the grating-mediated coupling dramatically complicates the landscape of photon generation.
In the presence of a grating, the total state can be represented as a superposition of converted and un-converted photon pairs according to Eq.~\eqref{JSA_total}. Since the two gratings work independently, they introduce four distinct combinations of OAM charges among the generated pairs, represented by the four terms --- each with their own phase-matching conditions --- in Eq.~\eqref{JSA_total}, compared to only one in the absence of a grating (Eq.~\eqref{eq:state_0}). We can thus write down the JSAs for each photon pair. In particular, for a pair composed of two photons with high OAM numbers $m_s'$ and $m_i'$, Eq.~\eqref{jsa.oam2} leads to
\begin{widetext}
\begin{align}\label{jsa.oam_2}
 \Phi^{m_s';-m_i'}(\omega_s,\bar{\omega}_i)
 =  \ &I^{m_s;-m_i}\,
     \int_0^L \mathrm{d}z \ e^{i\left(k_1-k_2- k_s^{m_s}
      + k_i^{-m_i} \right)z}
      \ \left[a_{m_s\rightarrow m_s'} (z) \right]^*
      \left[a_{-m_i' \leftarrow -m_i} (z)\right]^* \nn\\
 &+      I^{m_s';-m_i}\,
     \int_0^L \mathrm{d}z \ e^{i\left(k_1-k_2- k_s^{m'_s}
      + k_i^{-m_i} \right)z}
      \ \left[a_{m'_s\rightarrow m_s'} (z) \right]^*
      \left[a_{-m_i' \leftarrow -m_i} (z)\right]^* \nn\\
       &+      I^{m_s;-m'_i}\,
     \int_0^L \mathrm{d}z \ e^{i\left(k_1-k_2- k_s^{m_s}
      + k_i^{-m'_i} \right)z}
      \ \left[a_{m_s\rightarrow m_s'} (z) \right]^*
      \left[a_{-m_i' \leftarrow -m'_i} (z)\right]^* \nn\\
       &+      I^{m_s';-m_i'}\,
     \int_0^L \mathrm{d}z \ e^{i\left(k_1-k_2- k_s^{m'_s}
      + k_i^{-m'_i} \right)z}
      \ \left[a_{m'_s\rightarrow m_s'} (z) \right]^*
      \left[a_{-m_i' \leftarrow -m'_i} (z)\right]^*.
      \end{align}
\end{widetext}

We can now embrace the result discussed in Fig.~\ref{fig:JSI_without_grating}, and note that FWM process favors the local generation of photons with \textit{low OAM orders}. This is due to the fact that the OAM mode overlap integral, see Eq.~\eqref{eq:IOO}, is greater for generated OAM modes with lower orbital numbers; i.e.~$|I^{m_s;-m_i}| \gg |I^{m'_s;-m_i}|, |I^{m_s;-m'_i}|, |I^{m'_s;-m'_i}|$ if $m'_s,\,m'_i\gg m_s,\,m_i$. This observation allows us to approximate the above JSA~\eqref{jsa.oam_2} by keeping only the summand in the first row that scales with the largest transverse overlap integral $I^{m_s;-m_i}$:
\begin{align}\label{jsa.oam_approx}
     \Phi^{m_s';-m_i'}(\omega_s,\bar{\omega}_i)
       \approx
      \,&I^{m_s;-m_i}\,
     \int_0^L \mathrm{d}z \ e^{i\left(k_1-k_2- k_s^{m_s}
      + k_i^{-m_i} \right)z} \nonumber \\
      &\times \left[a_{m_s\rightarrow m_s'} (z) \right]^*
      \left[a_{-m_i' \leftarrow -m_i} (z)\right]^*.
\end{align}
Similarly approximated JSAs (introduced as part of the generated state in Eq.~\eqref{JSA_total}) corresponding to other pairs of modes are given in Appendix~\ref{sec:OAM_generation}.

Before we present the results of numerical calculations, it is worth considering how the gratings modify the phase-matching condition in the integral in Eq.~\eqref{jsa.oam_approx}. In the idealized case in which the transmission gratings are individually chosen to perfectly meet the frequency- ($\delta_{s/i}=0$) and phase-matching ($d_{s/i}=0$) criteria, the JSI in Eq.~\eqref{jsa.oam_approx} simplifies to
\begin{align}\label{eq:simplified_JSA}
 &\Phi^{m_s';-m_i'}(\omega_s,\bar{\omega}_i) \nn\\
 &= I^{m_s;-m_i}
     \int_0^L \mathrm{d}z \;  e^{i\left(k_1- k_2- k_s^{m_s}
      + k_i^{-m_i} \right)z}\nn\\
&\qquad\qquad\qquad\qquad\qquad\times\sin\left[\gamma_s(L-z)\right]\sin\qty(\gamma_i z).
\end{align}
By expressing the trigonometric functions in the above formulation as differences of exponentials, we find that the two gratings split each integral of type given in Eq.~\eqref{eq:simplified_JSA} into four different terms.

The amplitudes of the individual terms listed in Eq.~\eqref{JSA_total} are dictated by the strength of the FWM processes (expressed through $I^{m_s;-m_i}$), the angular momentum conservation (ensured by $\kappa_i$ and $\kappa_s$), and phase-matching of the grating-mediated conversion process. Using this last characteristic, we can therefore promote, or suppress different OAM conversion processes, by choosing the periods of the signal (idler) grating to be either resonant, or non-resonant with $k_s^{m_s'}-k_s^{m_s}$ ($k_i^{m_i'}-k_i^{m_i}$).

\begin{figure}[ht!]
 \includegraphics[width=.9\columnwidth]{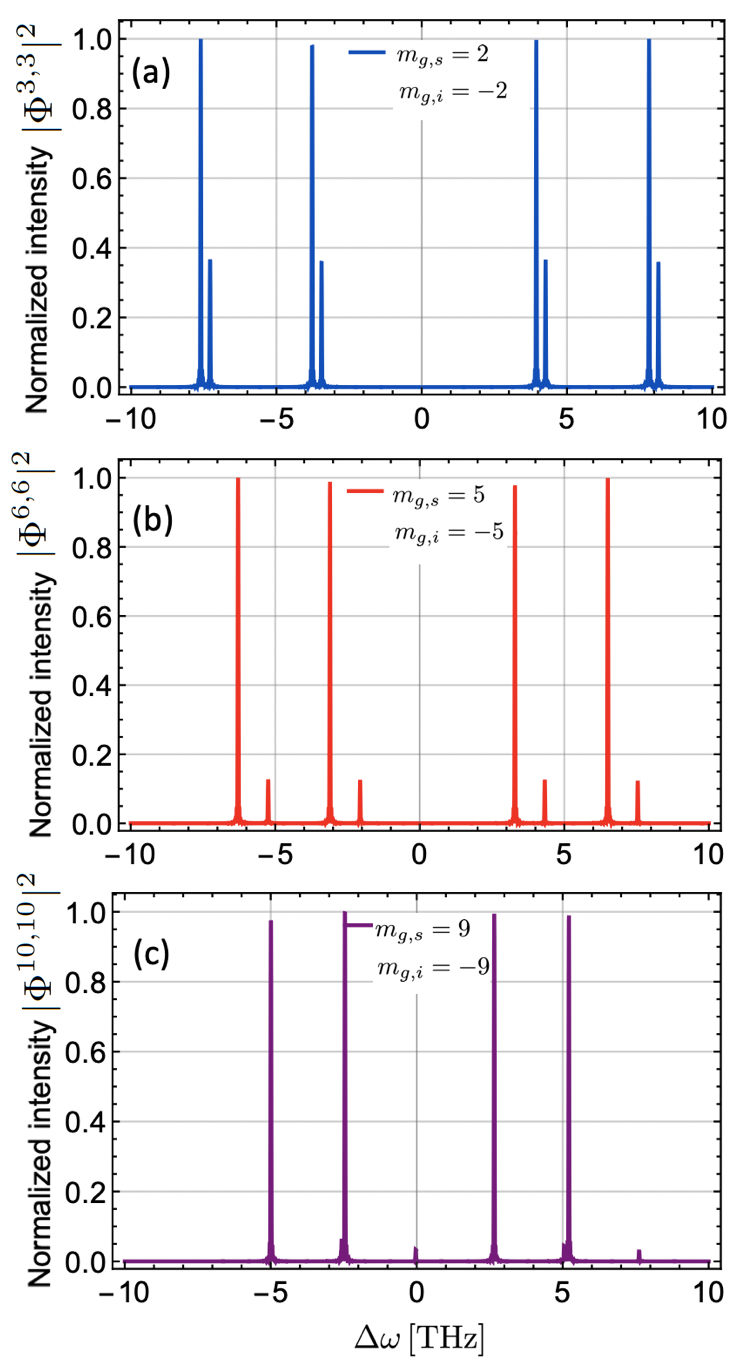}
  \caption{Normalized JSIs $\qty|\Phi^{m'_s,-m'_i}|^2\Big/\qty|\Phi_\tx{max}^{m'_s,-m'_i}|^2$ describing the generation of pairs of OAM modes $\mathbf{O}_{+m'_s}^+$ and $\mathbf{O}_{-m'_i}^-$ in the presence of gratings with topological charges (a) $m_{g,s}=-m_{g,i}=2$, (b) $m_{g,s}=-m_{g,i}=5$ and (c) $m_{g,s}=-m_{g,i}=9$, which are set in
  resonance with the OAM modes $\mathbf{O}_{+1}^+$ and $\mathbf{O}_{-1}^-$, respectively.  $\Delta \omega = \omega_s - \omega_{1} = \omega_{2} - \bar{\omega}_i$ is the frequency detuning of the modes, see Fig.~\ref{fig:grating}(a). }
 \label{fig:new_res_JSIs}
\end{figure}

In Fig.~\ref{fig:new_res_JSIs}, we show the numerical results for the JSIs $\qty|\Phi^{m'_s,-m'_i}|^2$ (the JSI of OAM-upconverted photon pairs) calculated for different gratings: (a) $m_{g,s}=-m_{g,i}=2$, (b) $m_{g,s}=-m_{g,i}=5$ and (c) $m_{g,s}=-m_{g,i}=9$, normalized in each case to the maximum value of the JSI, and plotted as a function of the detuning $\Delta\omega=\omega_s-\omega_1$.  In these plots we explicitly account only for the conversion from the $\mathbf{O}_{1}^+$ and $\mathbf{O}_{-1}^-$ pairs to $\mathbf{O}_{m_{g,s}+1}^+$ (so that $m_{s}'=m_{g,s}+1$) and $\mathbf{O}_{-(m_{g,i}-1)}^-$ (so that $m_{i}'=m_{g,i}-1$) --- this selectivity can be implemented in a physical system by choosing the grating periods to resonantly match that specific processes, while suppressing others.

For each grating setup, we identify eight features: the four strongest ones correspond to the dominant, first line in Eq.~\eqref{jsa.oam_2}, characterized by the strongest FWM, or the largest overlap integral. The second and third lines of Eq.~\eqref{jsa.oam_2} hardly contribute to the final JSI due to the very small overlap integrals, see Tables~\ref{T2}-\ref{T8}.  The far weaker peaks are due to the last term in Eq.~\eqref{jsa.oam_2}. As we increase the topological charge of the grating from (a) to (c), these weaker peaks become strongly suppressed justifying the approximation made in  Eq.~\eqref{jsa.oam_approx}.

\begin{figure}[ht!]
 \includegraphics[width=.95\columnwidth]{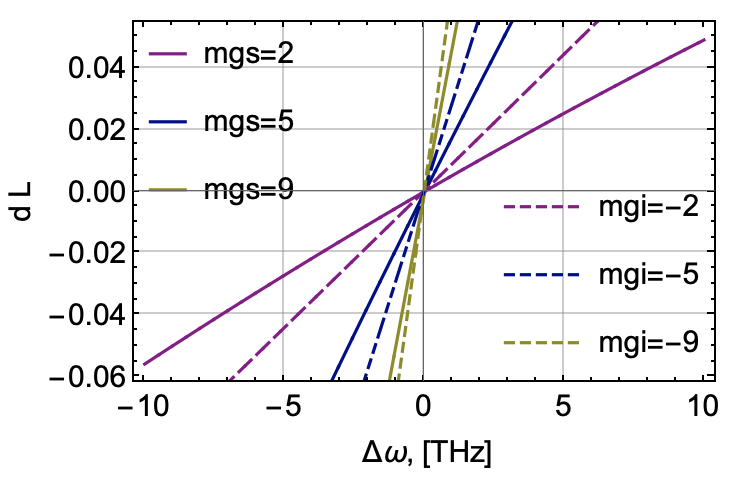}
        \caption{(a) Momentum detunings $d_{s/i}$ multiplied with the length $L$ of the waveguide for the resonance conversion shown in Fig.~\ref{fig:new_res_JSIs}  versus the frequency detuning $\Delta\omega$ for different grating orders $m_{g,s}$ and $m_{g,i}$ of the signal and idler modes, respectievely. }
    \label{fig:d_resonance}
\end{figure}

\begin{figure}[ht!]
 \includegraphics[width=.95\columnwidth]{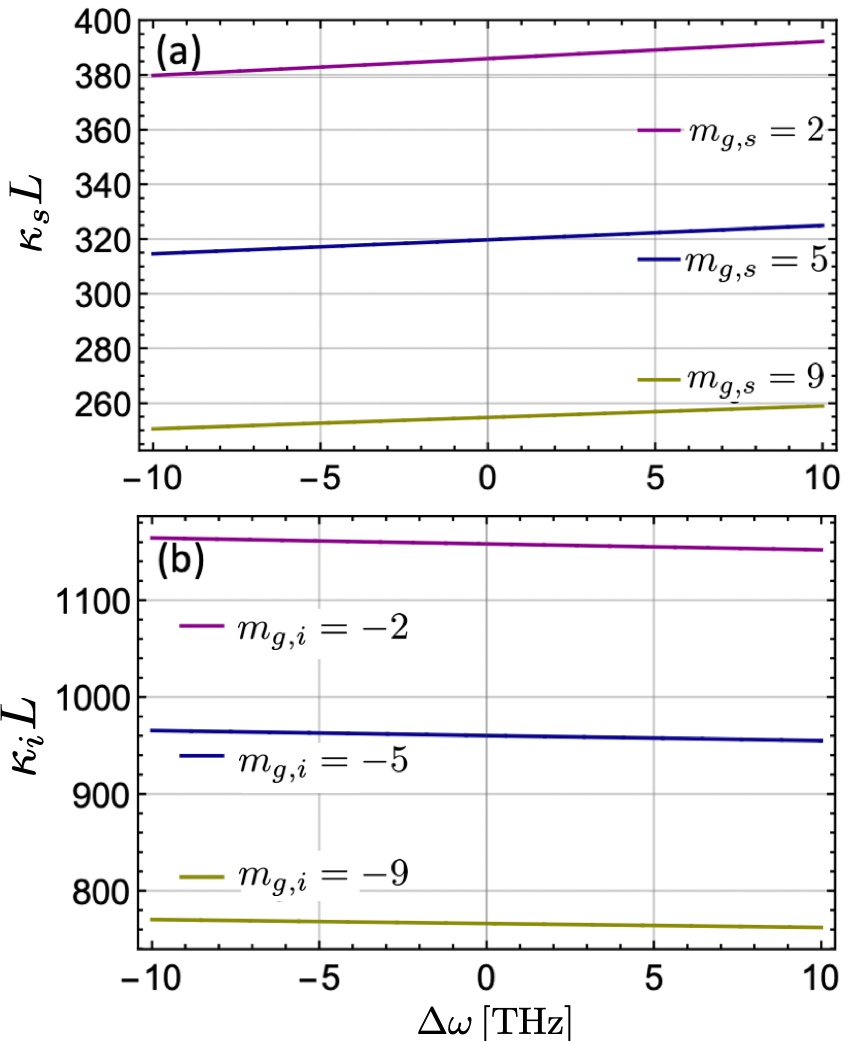}
        \caption{(a) Coupling constants between the input mode $\mathbf{O}_{+1}^+$ and the converted mode $\mathbf{O}_{m_{g,s}+1}^+$ multiplied with the length $L$ of the waveguide versus the frequency detuning $\Delta\omega$ for different grating orders $m_{g,s}$ of the signal mode. (b) The same for the idler modes $\mathbf{O}_{-1}^-$ and  $\mathbf{O}_{m_{g,i}-1}^-$, with negative grating numbers $m_{g,i}$.}
    \label{fig:kappa_resonance}
\end{figure}

Simultaneously, with  increasing OAM added by the gratings, all the peaks appear to be converging towards vanishing detuning $\Delta \omega$. To understand this effect, we note that the splitting between the modes is largely governed by the coupling parameters $\kappa_{s/i}$, which determine the amplitude of the JSA (see Eq.~\eqref{eq:simplified_JSA}). Indeed, the grating parameters are chosen to operate in the resonant regime, meaning that the momentum detunings $d_{s/i}=\frac{1}{2} \Big[\delta_s\qty(1/v_{g}^{m'_{s/i}}-1/v_{g}^{m_{s/i}})+K_{s/i}+k_{s/i}^{m_{s/i}}(\omega_{s/i})$ $-k_{s/i}^{m'_{s/i}}(\omega_{s/i})\Big]$ almost vanish for the chosen frequency range, see Fig.~\ref{fig:d_resonance}, and therefore $\gamma_{s/i} \approx \kappa_{s/i}$.
{Furthermore, as shown in Fig.~\ref{fig:kappa_resonance}, both (signal and idler) coupling parameters decrease with the increase in the topological charges $|m_{g,s/i}|$ for all frequency detunings $\Delta \omega$, leading to $\Delta \omega =0$ converging peaks to in Fig.~\ref{fig:new_res_JSIs} from (a) to (c).}

Additionally, it is noted that the profiles of the other (non-shown) JSIs $\qty|\Phi^{m'_s,-m_i}|^2$, $\qty|\Phi^{m_s,-m'_i}|^2 $ and $\qty|\Phi^{m_s,-m_i}|^2$ are identical to $\qty|\Phi^{m'_s,-m'_i}|^2$.

\begin{figure}[ht!]
 \includegraphics[width=\columnwidth]{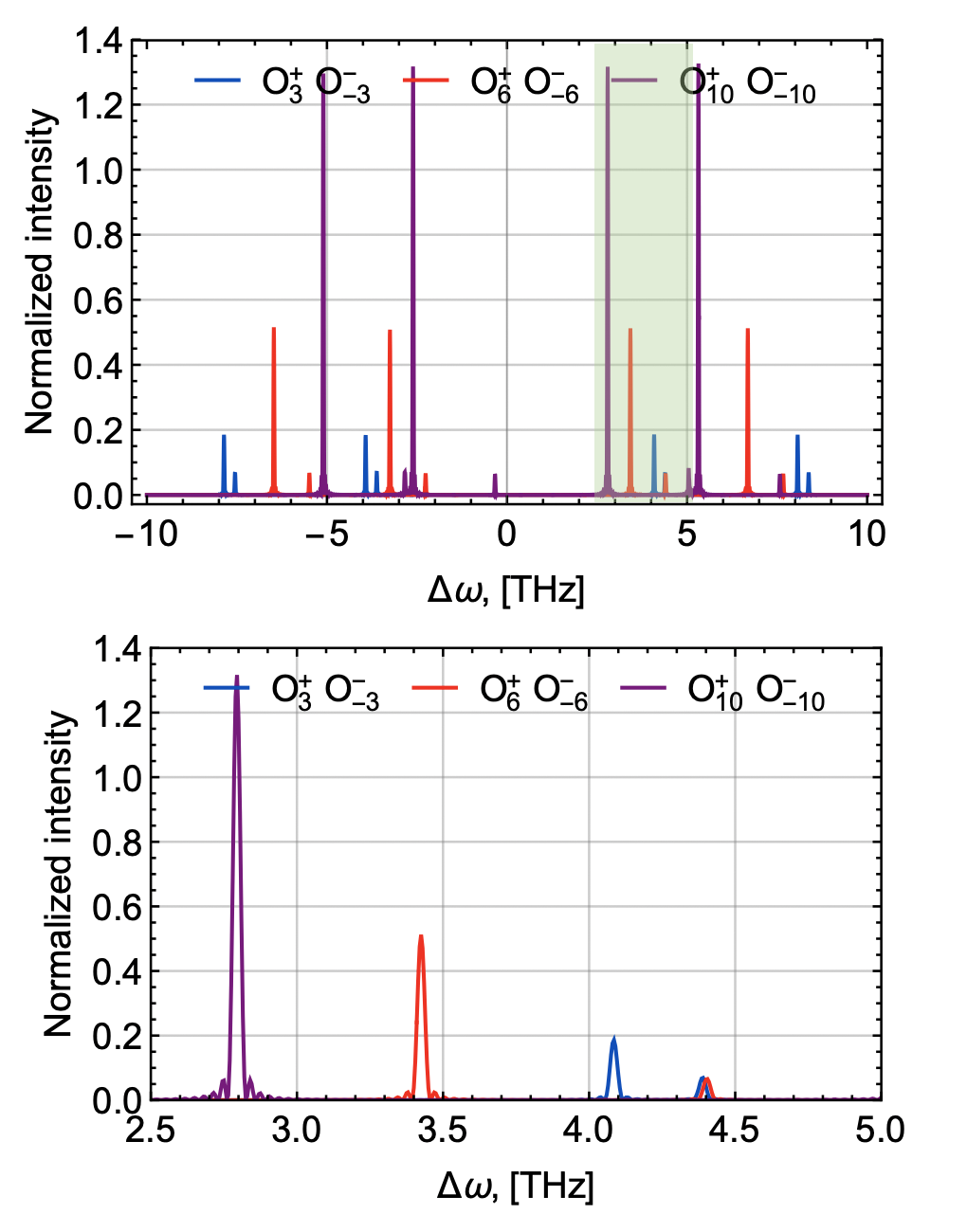}
        \caption{Normalized JSIs $\qty|\Phi^{m'_s,-m'_i}|^2\Big/\qty|\Phi_\tx{0,max}^{m'_s,-m'_i}|^2$ describing the generation of pairs of OAM modes $\mathbf{O}_{+m'_s}^+$ and $\mathbf{O}_{-m'_i}^-$ in the presence of gratings with topological charges $m_{g,s} = m'_s -1$ and $m_{g,i} = -m'_i +1$. The latter are in resonance with the OAM modes $\mathbf{O}_{+1}^+$ and $\mathbf{O}_{-1}^-$, respectively. The JSIs are normalized to the maximum JSIs $\qty|\Phi_\tx{0,max}^{m'_s,-m'_i}|^2$, which describe the generation of the same order OAM modes in a FWM process without a grating. The figure on the bottom is the zoom of the shaded green area in the figure on the top.
       }
    \label{fig:JSI_with_grating}
\end{figure}

{In Fig.~\ref{fig:JSI_with_grating} we show how much more efficient the FWM process in grating waveguides is for generating high-order OAM modes.
In this figure, the JSIs $\qty|\Phi^{m'_s,-m'_i}|^2$ describing the generation of pairs of OAM modes $\mathbf{O}_{+m'_s}^+$ and $\mathbf{O}_{-m'_i}^-$ in the presence of gratings with topological charges $m_{g,s}=m'_s-1$ and $m_{g,i}=m'_i+1$ which are set in resonance with the OAM modes $\mathbf{O}_{+1}^+$ and $\mathbf{O}_{-1}^-$, respectively, are normalized with respect to the maximum  JSI values corresponding to the generation of the same OAM modes, but in a waveguide without the grating  $\qty|\Phi_\tx{0, max}^{m'_s,-m'_i}|^2$.
As expected, due to the efficient conversion, the intensity of the generated OAM modes in the presence of a grating  grows  rapidly with increasing topological charge compared to the case without the grating.}

{The resulting relative efficiency given by the total number of photon pairs generated in the modes $\mathbf{O}_{m'_s}^{+}$ and $\mathbf{O}_{-m'_i}^{-}$, compared to the number of photon pairs generated in the waveguide without the grating,
\begin{equation}
	\frac{N^{ m'_s, -m'_i}_\mathrm{pairs}}{ N_0^{m_s,-m_i}}=
       \frac{ \int\mathrm{d} \omega_s \,
		\qty|\Phi^{m'_s,-m'_i}|^2}{ \int\mathrm{d} \omega_s \,
		\qty|\Phi_0^{m'_s,-m'_i}|^2},
\label{eq:Npairs}
\end{equation}
is presented in Tab.~\ref{tab:npairs}.
\begin{table}[ht]
\begingroup
\setlength{\tabcolsep}{10pt} 
\renewcommand{\arraystretch}{1.5} 
\begin{center}
	\begin{tabular}{c|c}
	Generated OAM modes & 	$N^{m_s,-m_i}_\mathrm{pairs}$ / $N_0^{m_s,-m_i}$ \\
   \hline
  $\mathbf{O}_{+3}^+$ $\mathbf{O}_{-3}^-$ &  1.05178   	 \\
		$\mathbf{O}_{+6}^+$ $\mathbf{O}_{-6}^-$           & 2.44514  \\
        $\mathbf{O}_{+10}^+$ $\mathbf{O}_{-10}^-$           & 7.03267 \\
	\end{tabular}
    \end{center}
	\caption{ Relative efficiency of grating waveguides for generating photon pairs carrying high-order OAM. 
 }
\label{tab:npairs}
\endgroup
\end{table}

This result clearly shows how grating waveguides allow us to enhance the generation efficiency of high-order OAM modes.

So far we have considered the \textit{resonant} processes, where the period of the signal (idler) gratings was chosen to selectively promote conversion between the $\mathbf{O}_{1}^+$ ($\mathbf{O}_{-1}^-$) and $\mathbf{O}_{m_{g,s}+1}^+$ ($\mathbf{O}_{-(m_{g,i}-1)}^-$) modes. To compare the efficiency of those resonant processes to that characterising non-resonant ones, Fig.~\ref{fig:non-resonance} shows the JSIs corresponding to the conversion of the locally generated modes $\mathbf{O}_{+2}^+$ and $\mathbf{O}_{-2}^-$ by the gratings with topological charges $m_{g,s}=9$ and $m_{g,i}=-9$ (assuming the grating chosen in resonance with the conversion from the $\mathbf{O}_{+1}^+$ and $\mathbf{O}_{-1}^-$ modes). The intensity profiles are normalized to the maximum of intensity of the resonant modes presented in Fig.~\ref{fig:new_res_JSIs}(c) {and shown by the dashed line in  Fig.~\ref{fig:non-resonance}.} We find that, unlike for the resonant case explored in Fig.~\ref{fig:new_res_JSIs}, these JSIs differ between different combinations of the emerging modes.

\begin{figure}[ht!]
 \includegraphics[width=.9\columnwidth]{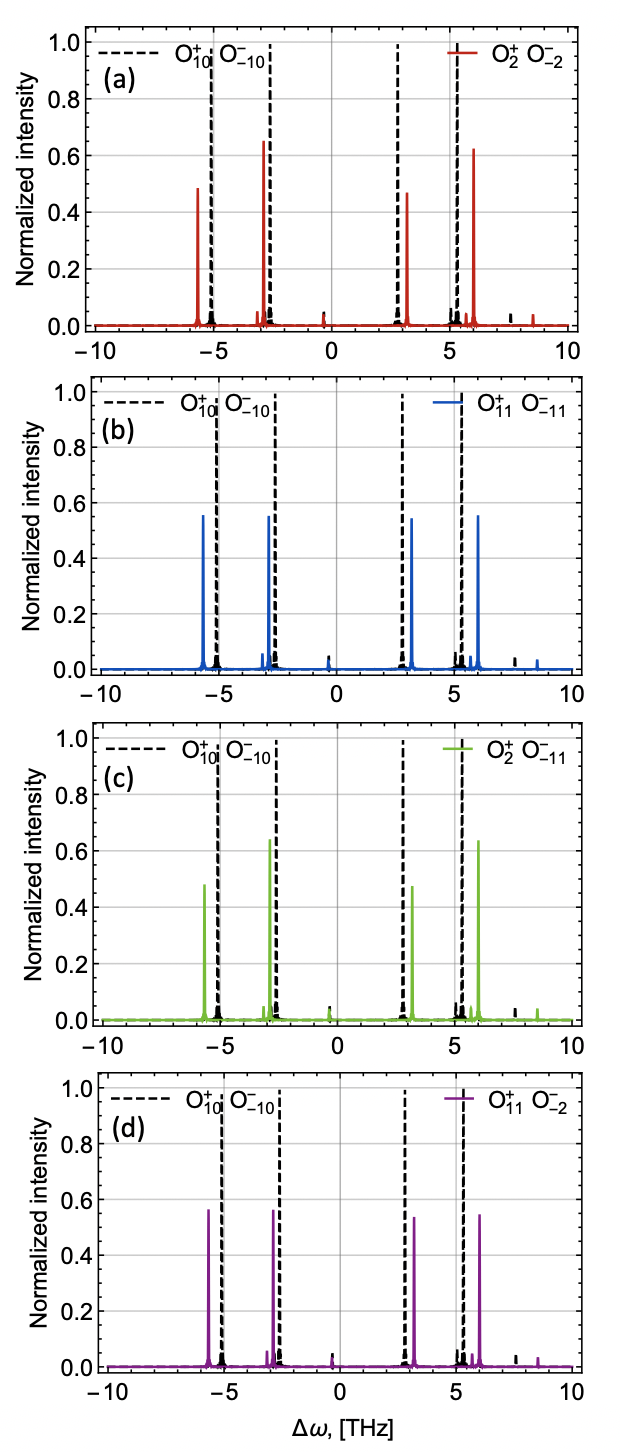}
        \caption{Conversion of the locally generated modes $\mathbf{O}_{+2}^+$ and $\mathbf{O}_{-2}^-$ by the gratings with  $m_{g,s}=9$ and  $m_{g,i}=-9$ (that results in the generation of $\mathbf{O}_{+2}^+$, $\mathbf{O}_{-2}^-$; $\mathbf{O}_{+2}^+$, $\mathbf{O}_{-11}^-$; $\mathbf{O}_{+11}^+$, $\mathbf{O}_{-2}^-$; $\mathbf{O}_{+11}^+$, $\mathbf{O}_{-11}^-$ sets of modes) which  are in resonance with the OAM modes $\mathbf{O}_{+1}^+$ and $\mathbf{O}_{-1}^-$. Presented JSIs are normalized with respect to  Fig.~\ref{fig:new_res_JSIs}
         (c) shown by the black dashed line in each plot.
         }
    \label{fig:non-resonance}
\end{figure}

Moreover, we also find that while the JSIs shown in Fig.~\ref{fig:non-resonance} are generally lower compared to the resonant modes, the effect of the suppression of non-resonant processes is not particularly strong. To understand this result, Fig.~\ref{fig:kappa_non-resonance} presents the coupling constants $\kappa_{s/i}$ and momentum detunings $d_{s/i}$, which make up the $\gamma_{s/i}$ parameters (see Eq.~\eqref{eq:param}) and enter the phase-matching conditions. We find that for the grating parameters used here, $\gamma_{s/i}$ is dominated by the $\kappa_{s/i}$, and largely constant across the entire considered spectral range. This flat response results in a weak selectivity of the resonant process. This selectivity could be improved by significantly lowering the strength of the grating, e.g.~by considering a smaller permittivity modulation amplitude $\Delta \varepsilon_0$, resulting in the stronger dependence of $\gamma_{s/i}$ on the detuning $d_{s/i}$. {At the same time, from Fig.~\ref{fig:non-resonance} it is clear that the frequencies at which the conversion of resonant and non-resonant modes occur are different, which makes it possible to filter resonantly converted modes using a frequency filter.}

\begin{figure}[ht!]
 \includegraphics[width=.9\columnwidth]{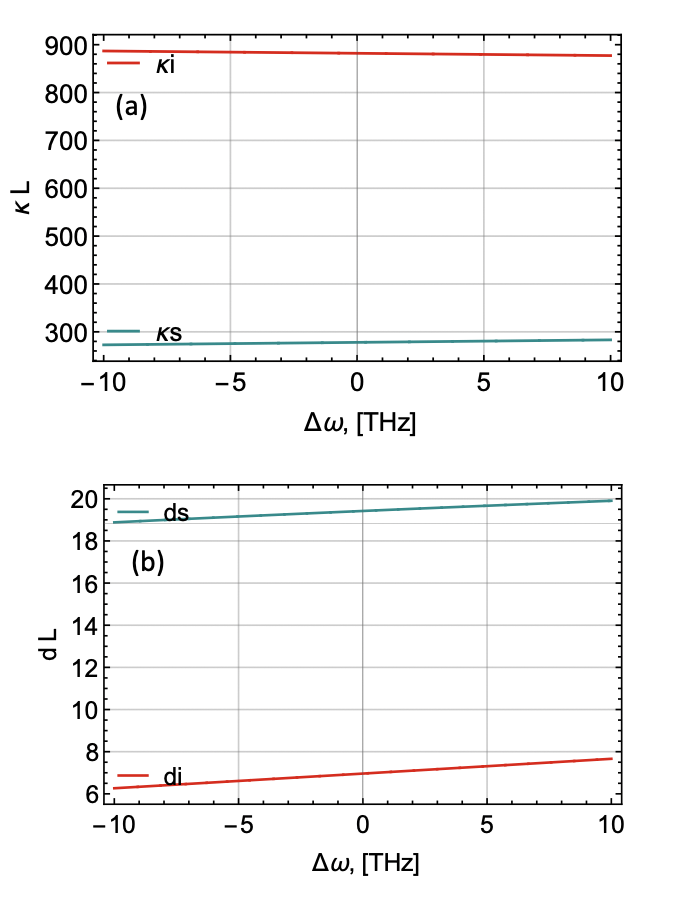}
        \caption{(a) Coupling constants between the input non-resonant signal (idler) mode  $\mathbf{O}_{+2}^+$ ($\mathbf{O}_{-2}^-$) and the converted mode $\mathbf{O}_{m_{g,s}+2}^+$ ($\mathbf{O}_{m_{g,i}-2}^-$) multiplied with the length of the waveguide versus frequency detuning. The topological charges are $m_{g,s}=9$, $m_{g,i}=-9$, that corresponds to the case in  Fig.~\ref{fig:non-resonance}. (b) Signal and idler detunings for the same case. }
    \label{fig:kappa_non-resonance}
\end{figure}

\subsection{Polarization entanglement}\label{subsec:entanglement}

Finally, the considered process leads to the generation of polarization entangled photons. Indeed,  the output state Eq.~\eqref{JSA_total} with the use of the approximation made in Eq.~\eqref{jsa.oam_approx} can be  written as

\begin{align}
    \ket{\psi_{\tx{gen}}} \approx 
         \int\mathrm{d}\omega_s&\;\Gamma(\omega_s,\bar{\omega}_i) \;
         \sum_{\sigma_s,\sigma_i=\pm} \ \  \sum_{m_s,m_i>0}\;
       I^{m_s, \sigma_s;-m_i, \sigma_i}  \nn\\
    		\times\Bigg\{ & \Phi_1
            \qty[\hat b_{s}^{\tx{O}_{m_s}^{\sigma_s}}(\omega_s)]^{\!\dagger} \;
            \qty[\hat b_{i}^{\tx{O}_{-m_i}^{\sigma_i}}(\bar{\omega}_i)]^{\!\dagger}  \nn\\
    		&+
            \Phi_2
            \qty[\hat b_{s}^{\tx{O}_{m_s}^{\sigma_s}}(\omega_s)]^{\!\dagger} \;
            \qty[\hat b_{i}^{\tx{O}_{-m_i'}^{\sigma_i}}(\bar{\omega}_i)]^{\!\dagger}
            \nn\\
    		 &+\Phi_3
            \qty[\hat b_{s}^{\tx{O}_{m_s'}^{\sigma_s}}(\omega_s)]^{\!\dagger} \;
            \qty[\hat b_{i}^{\tx{O}_{-m_i}^{\sigma_i}}(\bar{\omega}_i)]^{\!\dagger}  \nn\\
            &+
    	\Phi_4
            \qty[\hat b_{s}^{\tx{O}_{m_s'}^{\sigma_s}}(\omega_s)]^{\!\dagger}\;
            \qty[\hat b_{i}^{\tx{O}_{-m_i'}^{\sigma_i}}(\bar{\omega}_i)]^{\!\dagger}\;
        \Bigg\}
            \, \vert 0 \rangle_s \vert 0 \rangle_i,
        \label{eq:JSA_ent_1}
\end{align}
where the two-photon amplitudes  $\Phi_1$, $\Phi_2$, $\Phi_3$ and $\Phi_4$ are given by Eqs.~\eqref{jsa.oam_approx},~\eqref{jsa.oam3},~\eqref{jsa.oam4} and~\eqref{jsa.oam5}, respectively.
For example, in the idealized case ($\delta_{s/i}=0$, $d_{s/i}=0$), these two-photon amplitudes
read:
\begin{subequations}
\begin{align}
\Phi_1=  \int_0^L \mathrm{d}z \;  e^{i\Delta k z}
\cos\left[\kappa_s(L-z)\right]\cos\qty(\kappa_i z), \\
\Phi_2=  \int_0^L \mathrm{d}z \;  e^{i\Delta k z}
\cos\left[\kappa_s(L-z)\right]\sin\qty(\kappa_i z), \\
\Phi_3=  \int_0^L \mathrm{d}z \;  e^{i\Delta k z}
\sin\left[\kappa_s(L-z)\right]\cos\qty(\kappa_i z), \\
\Phi_4=  \int_0^L \mathrm{d}z \;  e^{i\Delta k z}
\sin\left[\kappa_s(L-z)\right]\sin\qty(\kappa_i z),
\label{eq:F_4}
\end{align}\end{subequations}
where $\Delta k = k_1- k_2- k_s^{m_s} + k_i^{-m_i}$. Expanding trigonometric functions in the expression above and integrating, one can notice that each function splits into four separate peaks (see Fig.~\ref{fig:new_res_JSIs}(c)) due to strong coupling (see Fig.~\ref{fig:kappa_resonance}).
 In the frequency space, the positions of these peaks are defined by the equation $\Delta k \mp \kappa_s \pm \kappa_i =0 $.
If the gratings are set in resonance with the modes $\mathbf{O}_{m_{s}}$ and $\mathbf{O}_{-m_{i}}$, the positions of the JSI peaks of resonant and non-resonant modes generated in such a process are shifted with respect to each other, see Fig.~\ref{fig:non-resonance}.
Therefore, by providing narrowband frequency filtering, it is possible to filter out one of the four components generated in the resonance process from the other three resonance components and from all non-resonance components.

Indeed, considering two filters in the form of a Dirac delta-function $\delta(\omega_s - (\omega_1+\Delta{\omega}))$ separately for the forward propagating signal photon and the backward propagating idler photon, where $\Delta{\omega}$ corresponds to the position of the desired peak, the generated state in Eq.~\eqref{eq:JSA_ent_1}
can be rewritten as
\begin{align}
    \ket{\psi_{\tx{gen}}} \approx
        e^{i \phi} \ \Gamma&(\omega_1+\Delta{\omega}) \;
        \left\{
    		I^{m_s,+;-m_i,-}\;\qty[\mathcal{A}^+_{m_s}]^{\!\dagger}\qty[\mathcal{B}^-_{-m_i}]^{\!\dagger} \right. \nonumber \\
                &\left.+\;I^{m_s,-;-m_i,+}\;\qty[\mathcal{A}^-_{m_s}]^{\!\dagger}\qty[\mathcal{B}^+_{-m_i}]^{\!\dagger} \right\}
            \vert 0 \rangle_s \vert 0 \rangle_i,
        \label{Bell}
\end{align}
where $\phi = \pm \kappa_s L /2$ depending on the filtered peak, and a set of supermodes is introduced:
\begin{subequations}
\begin{align}
\qty[\mathcal{A}^{\pm}_{m_s}]^{\!\dagger}\!= &\qty[\hat b_{s}^{\tx{O}_{m_s}^{\pm}}( \omega_1+\Delta{\omega}))]^{\!\dagger} \! +  \qty[\hat b_{s}^{\tx{O}_{m_s+m_{g,s}}^{\pm}}( \omega_1+\Delta{\omega}))]^{\!\dagger}\!, \\
\qty[\mathcal{B}^{\mp}_{-m_i}]^{\!\dagger}\!= &\qty[\hat b_{i}^{\tx{O}_{-m_i}^{\mp}}(\omega_2-\Delta{\omega}))]^{\!\dagger}
\!+  \qty[\hat b_{i}^{\tx{O}_{-m_i+m_{g,i}}^{\mp}}(\omega_2-\Delta{\omega})
]^{\!\dagger}\!.
  \label{supermodes}
\end{align}\end{subequations}
Note that in Eq.~\eqref{Bell} any spin-orbit coupling effects are neglected.
The state \eqref{Bell} exhibits the form of the polarization (spin) entangled Bell-state $\ket{\psi_{\tx{Bell}}} = \frac{1}{\sqrt{2}}(  \vert + \rangle_s \vert - \rangle_i + \vert - \rangle_s \vert + \rangle_i)$.

The fidelity of generating such a state is mainly determined by the difference of overlapping integrals $I^{m_s,+;-m_i,-}$ and $I^{m_s,-;-m_i,+}$, which are presented in the Appendix~\ref{sec:overlapintegral}.  Then, according to Tab.~\ref{T2} and Tab.~\ref{T6} of the Appendix, the fidelity of generating polarization entangled states, $F=\braket{\psi_\text{Bell}}{\psi_\text{gen}}$, for gratings which are set in resonance with the mode orders $m_s$ and $-m_i$ with the use of a delta-shaped filter, is given in Table~\ref{tab:fidelity}.


\begin{table}[ht]
	\centering
	\begin{tabular}{c|c|c|c|c}
		$-m_i / m_s$ &    1	 &    2   &    3   &    4   \\
  \hline
		1          & 0.999995 & 0 & 0.903565 & 0 \\
    \hline
        2           &0 & 0.999997 & 0 & 0.891517 \\
          \hline
        3           & 0.902705 & 0 & 0.999993 & 0 \\
          \hline
        4         & 0 & 0.891517 & 0& 0.999985\\
	\end{tabular}
	\caption{Fidelity $F=\braket{\psi_\text{Bell}}{\psi_\text{gen}}$ depending on the order of resonance mode.}
	\label{tab:fidelity}
\end{table}
One can notice that the fidelity (especially for the generated signal and idler photons with opposite OAM, see diagonal elements in Tab.\ref{tab:fidelity}) is close to one, showing a high efficiency of generating polarization entangled photons in the considered process.

\section{Conclusion}\label{sec:conclusion}

In this work, we  theoretically demonstrated an efficient generation of the  counterpropagating and polarization entangled photons in OAM modes with non-zero topological charges  during the FWM process in a silica fiber with two helical gratings. Topological charges and the helicity of the gratings can be varied to create desired topological charges of OAM modes. Due to opposite propagation directions, such entangled photons can be easily sorted. It was shown that the frequencies of the generated photons have a shift compared to the pump frequencies. Such a shift depends on the topological charge of the grating and allows a simple filtering of the OAM modes from the pump, which is impossible without the grating.  We believe that such polarization entangled photons with high-order OAM can be a promising tool for a storage and transferring quantum information, as well as for realization different quantum information protocols.

\section{Acknowledgement}
Financial support of the Deutsche Forschungsgemeinschaft(DFG) through the projects SH 1228/3-1 and SFB/TRR-142 (subproject C10) is gratefully acknowledged. MKS and MJS acknowledge funding from the Macquarie University Research Fellowship (MQRF0001036) and the Australian Research Council (Discovery Project DP160101691 and Discovery Early Career Researcher Award DE220101272).

\clearpage

\appendix

\section{Cylindrical dielectric waveguides}\label{sec:waveguides}

The radial field profiles for the HE/EH modes are~\cite{Okamoto}:
\begin{align}
    \mathbf{e}_{m,n}^\tx{HE/EH}(r)=\begin{cases}
      \begin{pmatrix}
          \frac{ik(\omega)}{u^2}\left[s\frac{m}{r}\bar J_m(r)-u\bar J'_m(r)\right]\\
          \frac{ik(\omega)}{u^2}\left[\frac{m}{r}\bar J_m(r)-s u\bar J'_m(r)\right]\\
          \bar J_m(r)
     \end{pmatrix}&\!\!\!\!\!\!\!\tx{, if }r\leq a\\
     \begin{pmatrix}
          -\frac{ik(\omega)}{w^2}\left[s\frac{m}{r}\bar K_m(r)-w\bar K'_m(r)\right]\\
          -\frac{ik(\omega)}{w^2}\left[\frac{m}{r}\bar K_m(r)-s w\bar K'_m(r)\right]\\
          \bar K_m(r)
    	\end{pmatrix}&\!\!\!\!\!\!\!\tx{, if }r> a
    \end{cases},
    \label{eq:radialpart}
\end{align}
with the normalized Bessel functions
\begin{subequations}\begin{align}
\bar J_m(r)&=\frac{J_m(ur)}{J_m(ua)},\quad \bar J'_m(r)&=\frac{J'_m(ur)}{J_m(ua)},\\
\bar K_m(r)&=\frac{K_m(wr)}{K_m(wa)},\quad \bar K'_m(r)&=\frac{K'_m(wr)}{K_m(wa)}
\end{align}\end{subequations}
and
\begin{align}
	s=\frac{m\left(\frac{1}{U^2}+\frac{1}{W^2}\right)}{\frac{J'_m(U)}{U J_m(U)}+\frac{K'_m(W)}{W K_m(W)}}
    \label{eq:s}.
\end{align}
normalized parameters $U=ua$, $W=wa$, and
\begin{align}
	V=\sqrt{U^2+W^2}
    =ka\sqrt{\varepsilon_\tx{co}-\varepsilon_\tx{cl}}
    =\frac{2\pi a}{\lambda}\sqrt{\varepsilon_\tx{co}-\varepsilon_\tx{cl}},
    \label{eq:V}
\end{align}
are defined using the lateral wavenumber $u$  and lateral attenuation constant $w$
\begin{subequations}
\begin{align}
	u\,r &= \sqrt{k_0^2\,\varepsilon_\tx{co}-k^2(\omega)}\ r \ \tx{ for }r<a,\\
    i\,w\,r &= i\sqrt{k^2(\omega)-k_0^2\,\varepsilon_\tx{cl}}\ r \ \tx{ for }r>a.
\end{align}\label{eq:uw}\end{subequations}

\section{Mode overlap integrals}\label{sec:overlapintegral}

The strength of the FWM process, which describes the generation of signal and idler modes exhibiting mode profiles in the OAM basis, is given by the transverse mode overlap integral in Eq.~\eqref{eq:IOO}:
\begin{align}
     &I^{\pm m_s,\sigma_s;\pm m_i,\sigma_i} = \sum_{\mu\nu\tau\rho}\;\chi^{(3)}_{\mu\nu\tau\rho}
        \iint \tx{d}\mathbf{r}\nn\\
       & \ \ \  \times \qty[
        	\left(\modeHE{HE}{1}{\ee}\right)_{\!\mu}
            \left(\modeHE{HE}{1}{\ee}\right)_{\!\nu}
            \left(\modeO{\pm m_s}{\sigma_s}\,\right)_{\!\tau}^{\!*}
            \left(\modeO{\pm m_i}{\sigma_i}\,\right)_{\!\rho}^{\!*}],
        \label{eq:OAMintegral}
\end{align}
where the indices $\mu,\nu,\tau,\rho\in\{r,\phi,z\}$ label in this notation the matrix elements of $\chi^{(3)}$, and the vector components of the respective modes in cylindrical coordinates. The integral ranges across the whole transverse plane of the waveguide, explicitly with limits $\int_0^\infty\tx{d}r\int_0^{2\pi}\tx{d}\phi$, where the theoretical infinite upper limit of the radial integral is numerically approximated in subsection~\ref{sec:overlapintergalresults}. The azimuthal integral, on the other hand, can be analytically derived as shown next.

\subsection{Azimuthal integral}\label{sec:azimuthalintergal}
In evaluating the azimuthal integral, one has to take only those parts of the electric field into account that vary with the azimuthal angle $\phi$. While all vector elements of the OAM modes vary identically with $\phi$ (see Eqs.~\eqref{eq:Oplusdef} and~\eqref{eq:Ominusdef},) the (here even) hybrid mode elements are described by either a $\cos \phi$ or a $\sin \phi$ function, see Eq.~\eqref{eq:evenhybrid}, such that one has to distinguish between the different vector components $\mu,\,\nu\in\{r,\phi,z\}$:
\begin{widetext}
\begin{align}
        &\int_0^{2\pi} \tx{d}\phi \qty[
        	\qty(\modeHE{HE}{1}{\ee})_{\!\mu}
            \qty(\modeHE{HE}{1}{\ee})_{\!\nu}
            \qty(\modeO{\pm m_s}{\sigma_s})_{\!\tau}^{\!*}
            \qty(\modeO{\pm m_i}{\sigma_i})_{\!\rho}^{\!*}] \nn\\
        &\propto\int_0^{2\pi} \tx{d}\phi \qty[
        	\qty(\mathbf{g}_m^{(\tx e)})_{\!\mu}
            \qty(\mathbf{g}_m^{(\tx e)})_{\!\nu}
            \qty(e^{i(\pm m_s+\tilde\sigma_s)\phi})^{\!*}
            \qty(e^{i(\pm m_i+\tilde\sigma_i)\phi})^{\!*}] \nn\\
        &=\begin{cases}
        \int_0^{2\pi} \tx{d}\phi \qty[
        	\cos\phi\cos\phi \, 
            e^{-i(\pm m_s+\tilde\sigma_s)\phi} \,
            e^{-i(\pm m_i+\tilde\sigma_i)\phi}], & \quad \tx{if } \mu,\nu\in\{r,z\} \\
        \int_0^{2\pi} \tx{d}\phi \qty[
        	 \sin\phi \sin\phi \,
             e^{-i(\pm m_s+\tilde\sigma_s)\phi}\,
             e^{-i(\pm m_i+\tilde\sigma_i)\phi}], & \quad \tx{if } \mu=\nu=\phi \\
        \int_0^{2\pi} \tx{d}\phi \qty[
        	\cos\phi \sin\phi \,
            e^{-i(\pm m_s+\tilde\sigma_s)\phi} \,
            e^{-i(\pm m_i+\tilde\sigma_i)\phi}], & \quad \tx{if } (\mu\in\{r,z\} \land \nu=\phi) \lor (\nu\in\{r,z\} \land \mu=\phi)
        \end{cases}\nn\\
        &=\begin{cases}
        \int_0^{2\pi} \tx{d}\phi \qty[
        	\cos^2\phi \,
            e^{-i(\pm m_s\pm m_i+\tilde\sigma_s+\tilde\sigma_i)\phi}], & \quad\quad \tx{if } \mu,\nu\in\{r,z\} \\
        \int_0^{2\pi} \tx{d}\phi \qty[
        	\sin^2\phi \,
            e^{-i(\pm m_s\pm m_i+\tilde\sigma_s+\tilde\sigma_i)\phi}], & \quad\quad \tx{if } \mu=\nu=\phi
 \\
        \int_0^{2\pi} \tx{d}\phi \qty[
        	\cos\phi\sin\phi\,
            e^{-i(\pm m_s\pm m_i+\tilde\sigma_s+\tilde\sigma_i)\phi}], & \quad\quad \tx{if } (\mu\in\{r,z\} \land \nu=\phi) \lor (\nu\in\{r,z\} \land \mu=\phi)
        \end{cases}\nn\\
        &=\begin{cases}
        \frac{1}{4}\int_0^{2\pi} \tx{d}\phi
            \qty[e^{-i(\pm m_s\pm m_i+\tilde\sigma_s+\tilde\sigma_i-2)\phi}
            +
            e^{-i(\pm m_s\pm m_i+\tilde\sigma_s+\tilde\sigma_i+2)\phi}
            +
            2e^{-i(\pm m_s\pm m_i+\tilde\sigma_s+\tilde\sigma_i)\phi}],
            &\!\!\!\!\!\!\!\!\!\!\!\!\!\!\!\!\!\!\!\!\!\!\!\!\!\!\!\!\!\!\!\!\!\!\!\!\!\!\!\!\!\!\!\! \tx{if } \mu,\nu\in\{r,z\}
            \\
            -\frac{1}{4}\int_0^{2\pi} \tx{d}\phi
            \qty[e^{-i(\pm m_s\pm m_i+\tilde\sigma_s+\tilde\sigma_i-2)\phi}
            +
            e^{-i(\pm m_s\pm m_i+\tilde\sigma_s+\tilde\sigma_i+2)\phi}
            -
            2e^{-i(\pm m_s\pm m_i+\tilde\sigma_s+\tilde\sigma_i)\phi}] , &\!\!\!\!\!\!\!\!\!\!\!\!\!\!\!\!\!\!\!\!\!\!\!\!\!\!\!\!\!\!\!\!\!\!\!\!\!\!\!\!\!\!\!\! \tx{if } \mu=\nu=\phi
\\
            \frac{1}{4i}\int_0^{2\pi} \tx{d}\phi
            \qty[e^{-i(\pm m_s\pm m_i+\tilde\sigma_s+\tilde\sigma_i-2)\phi}
            -
            e^{-i(\pm m_s\pm m_i+\tilde\sigma_s+\tilde\sigma_i+2)\phi}], \ \ \tx{if } (\mu\in\{r,z\} \land \nu=\phi) \lor (\nu\in\{r,z\} \land \mu=\phi)
        \end{cases}
        \label{eq:intphi}
\end{align}
\end{widetext}
where in the last step the following identities are used:
\begin{align}
    \cos^2\phi &= \qty(\,\frac{1}{2}\,\qty(e^{i\phi}+e^{-i\phi}))^{\!2}
    = \frac{1}{4}\qty(e^{2i\phi}+e^{-2i\phi}+2), \nn\\
    \sin^2\phi &= \qty(\frac{1}{2i}\qty(e^{i\phi}-e^{-i\phi}))^{\!2}
    =-\frac{1}{4}\qty(e^{2i\phi}+e^{-2i\phi}-2), \nn\\
    (\cos\phi)&(\sin\phi) = \frac{1}{4i}\qty(e^{2i\phi}-e^{-2i\phi}).
\end{align}
The integrals in Eq.~\eqref{eq:intphi} are all vanishing iff the arguments of the Euler's numbers are non-zero, i.e.~in general $\int_0^{2\pi} e^{im\phi}=0\ \forall \ m\in \mathbb{Z}/\{0\}$, due to the periodicity of $e$ (or Euler's identity). Thus, the total angular momentum conservation rule is recovered:
\begin{align}\label{eq:angmomcond}
    \pm m_s \pm m_i + \tilde\sigma_s + \tilde\sigma_i \in \{\pm 2,\,0\},
\end{align}
which has to be satisfied during the FWM process.

\subsection{Note on the \texorpdfstring{$\chi^{(3)}$}{chi3} tensor}\label{sec:noteonchi3}
Next, it is to note that for calculating the total transverse integral~\eqref{eq:OAMintegral}, its integrand can be written explicitly in terms of the $\chi^{(3)}$ tensor --- assuming the simplified, homogeneous and isotropic form\cite{boyd}
\begin{equation}
    \chi^{(3)} = \chi_0^{(3)}
\begin{pmatrix}
1 & 1/3 & 1/3 & 0 & 0 & 0 \\
1/3 & 1 & 1/3 & 0 & 0 & 0 \\
1/3 & 1/3 & 1 & 0 & 0 & 0 \\
0 & 0 & 0 & 1/3 & 0 & 0 \\
0 & 0 & 0 & 0 & 1/3 & 0 \\
0 & 0 & 0 & 0 & 0 & 1/3
\end{pmatrix},
\label{eq:chi3}
\end{equation}
the product of the $\chi^{(3)}$ tensor with the four modes profiles gives:
\begin{align}
    &\chi^{(3)} \;
        	\underbrace{\modeHE{HE}{1}{\ee}}_{A} \;
            \underbrace{\modeHE{HE}{1}{\ee}}_{B} \;
            \underbrace{\qty(\modeO{\pm m_s}{\sigma_s})^{\!*}}_{C} \;
            \underbrace{\qty(\modeO{\mp m_i}{\sigma_i})^{\!*}}_{D} \nn\\
    &= \chi_0^{(3)} \Big(
    A_r B_r C_r D_r + A_\phi B_\phi C_\phi D_\phi + A_z B_z C_z D_z \nn\\
    &\quad\ + \frac{1}{3}[ A_r B_r C_\phi D_\phi + A_r B_r C_z D_z + A_\phi B_\phi C_z D_z \nn\\
    &\quad\ + A_\phi B_\phi C_r D_r + A_z B_z C_r D_r + A_z B_z C_\phi D_\phi \nn\\
    &\quad\  + A_\phi B_z C_\phi D_z + A_\phi B_z C_z D_\phi + A_z B_\phi C_\phi D_z \nn\\
    &\quad\ + A_z B_\phi C_z D_\phi+ A_r B_z C_r D_z + A_r B_z C_z D_r  \nn\\
    &\quad\  + A_z B_r C_r D_z + A_z B_r C_z D_r + A_r B_\phi C_r D_\phi \nn\\
    &\quad\ + A_r B_\phi C_\phi D_r + A_\phi B_r C_r D_\phi + A_\phi B_r C_\phi D_r ]
    \Big),
\end{align}
where the subscripts $(r,\phi,z)$ indicate, as previously, the field components in cylindrical coordinates.

\subsection{Results of the overlap integrals}\label{sec:overlapintergalresults}
Finally, the results of the overlap integral in Eq.~\eqref{eq:OAMintegral} are shown for different combinations of OAM modes in tables~\ref{T2} to~\ref{T8}. Note that all overlap integrals associated with the generation of signal and idler modes $\mathbf{O}_{m_s}^+ \mathbf{O}_{m_i}^+$ and $\mathbf{O}_{-m_s}^- \mathbf{O}_{-m_i}^-$ are zero, via \eqref{eq:angmomcond}.

\begin{table}[ht]
	\centering
	\begin{tabular}{c|c|c|c|c}
		$m_i / m_s$ &    1	 &    2   &    3   &    4   \\
  \hline
		1          & 0.225874 & 0 & -0.000112 & 0 \\
    \hline
        2           &0 & 0.164477 & 0 & -0.000091 \\
          \hline
        3           & -0.000112 & 0 & 0.124059 & 0 \\
          \hline
        4         & 0 & -0.000091 & 0& 0.093515\\
	\end{tabular}
	\caption{Overlap integrals~\eqref{eq:OAMintegral} for the generation of two OAM modes with co-rotation SAM and OAM: $\mathbf{O}_{m_s}^+ \mathbf{O}_{-m_i}^-$ and $\mathbf{O}_{-m_s}^- \mathbf{O}_{m_i}^+$.}
	\label{T2}
\end{table}

\begin{table}[ht]
	\centering
	\begin{tabular}{c|c|c|c|c}
		$m_i / m_s$ &    1	 &    2   &    3   &    4  \\
  \hline
		1          & 0.079653 & 0 & 0.001285 & 0  \\
    \hline
        2         & 0 & 0.082187 & 0 & 0.001038\\
          \hline
        3          &0 & 0 & 0.061948 & 0 \\
          \hline
        4           &0 & 0 & 0 & 0.046662\\
	\end{tabular}
	\caption{Overlap integrals~\eqref{eq:OAMintegral} for the generation of  $\mathbf{O}_{m_s}^+ \mathbf{O}_{m_i}^-$ and $\mathbf{O}_{-m_s}^- \mathbf{O}_{-m_i}^+$ pairs of modes. }
	\label{T3}
\end{table}

\begin{table}[ht]
	\centering
	\begin{tabular}{c|c|c|c|c}
		$m_i / m_s$ &    1	 &    2   &    3   &    4  \\
  \hline
		1        & 0.079653 &\ \ \  \ \ 0 \ \ \ \ \ & \ \  \ \ \ 0 \  \ \ \ \ & \ \ \ \  \ 0 \ \  \ \ \  \\
    \hline
        2         & 0 & 0 & 0 & 0\\
          \hline
        3          & 0 & 0 & 0 & 0 \\
          \hline
        4          & 0 & 0 & 0 & 0\\
	\end{tabular}
	\caption{Overlap integrals~\eqref{eq:OAMintegral} for the generation of  $\mathbf{O}_{m_s}^+ \mathbf{O}_{-m_i}^+$ and $\mathbf{O}_{-m_s}^- \mathbf{O}_{m_i}^-$ pairs of modes. }
	\label{T4}
\end{table}

\begin{table}[ht]
	\centering
	\begin{tabular}{c|c|c|c|c}
		$m_i / m_s$ &    1	 &    2   &    3   &    4  \\
  \hline
		1       &0.224501 & 0& 0.000314& 0  \\
    \hline
        2       & 0 & 0.163736 & 0 & 0.000279\\
          \hline
        3        & 0.000316& 0 & 0.123113 & 0 \\
          \hline
        4        & 0 & 0.000279 & 0 & 0.092491\\
	\end{tabular}
	\caption{Overlap integrals~\eqref{eq:OAMintegral} for the generation of  $\mathbf{O}_{-m_s}^+ \mathbf{O}_{m_i}^-$ and $\mathbf{O}_{m_s}^- \mathbf{O}_{-m_i}^+$ pairs of modes.}
	\label{T5}
\end{table}

\begin{table}[ht]
	\centering
	\begin{tabular}{c|c|c|c|c}
		$m_i / m_s$ &    1	 &    2   &    3   &    4  \\
  \hline
		1       & 0.079652 & 0 & 0 & 0 \\
    \hline
        2      & 0 & 0.082186 & 0 & 0\\
          \hline
        3        & 0.001291 & 0 & 0.061946 & 0 \\
          \hline
        4       & 0 & 0.001039 & 0 & 0.046660\\
	\end{tabular}
	\caption{Overlap integrals~\eqref{eq:OAMintegral} for the generation of  $\mathbf{O}_{-m_s}^+ \mathbf{O}_{-m_i}^-$ and $\mathbf{O}_{m_s}^- \mathbf{O}_{m_i}^+$ pairs of modes. }
	\label{T6}
\end{table}

\begin{table}[ht]
	\centering
	\begin{tabular}{c|c|c|c|c}
		$m_i / m_s$ &    1	 &    2   &    3   &    4  \\
  \hline
		1        & 0.079652 &\ \ \  \ \ 0 \ \  \ \ \ & \ \ \  \ \ 0 \  \ \ \ \ & \ \ \ \ \ 0 \ \ \  \ \  \\
    \hline
        2         & 0 & 0 & 0 & 0\\
          \hline
        3          & 0 & 0 & 0 & 0 \\
          \hline
        4          & 0 & 0 & 0 & 0\\
	\end{tabular}
	\caption{Overlap integrals~\eqref{eq:OAMintegral} for the generation of  $\mathbf{O}_{-m_s}^+ \mathbf{O}_{m_i}^+$ and $\mathbf{O}_{m_s}^- \mathbf{O}_{-m_i}^-$ pairs of modes. }
	\label{T7}
\end{table}

\begin{table}[ht]
	\centering
	\begin{tabular}{c|c|c|c|c}
		$m_i / m_s$ &    1	 &    2   &    3   &    4  \\
  \hline
		1      & 0.224501 & 0 & 0.000314 & \ \ \ \  \ 0 \ \ \ \ \ \\
    \hline
        2     & 0 & 0.000016 & 0 & 0\\
          \hline
        3        &0.000316 & 0 & 0 & 0  \\
          \hline
        4       &0 & 0 & 0 & 0\\
	\end{tabular}
	\caption{Overlap integrals~\eqref{eq:OAMintegral} for the generation of  $\mathbf{O}_{-m_s}^+ \mathbf{O}_{-m_i}^+$ and $\mathbf{O}_{m_s}^- \mathbf{O}_{m_i}^-$ pairs of modes. }
	\label{T8}
\end{table}

\newpage
\section{Asymptotic-in and -out modes}\label{sec:input_output}

The derivation of the generated state after the FWM process is presented assuming the more general ansatz that the mode profiles of the signal and idler photons exhibit not only the OAM mode profiles whose topological charges equal the number of spirals of the corresponding transmission grating, but of a superposition of the grating phase-matched OAM mode and the fundamental HE mode. For this, the asymptotic-in and -out state formalism described in~\cite{asymptotic} is used and the coupled mode equation for three interacting modes are assumed --- these are one mode from the pump photon, one mode from the HE part of the generated photon and one mode from the OAM part of the generated photon; for each the signal and idler photon.

\subsection{Signal photons}

In Section \ref{sec:fwm}, we have provided the explicit expansion of the electric field operator associated with the signal photon, listing the coupled mode equations and boundary conditions for the component describing the generation of an OAM photon with the momentum $m_s$ at the end of the waveguide, as well as presented the solution for the amplitudes $a_{m_s\rightarrow m_s}(z)$ and $a_{m'_s\rightarrow m_s}(z)$ in Eq.~\eqref{eq:as1}.

Similarly, we can define the amplitudes defining Eq.~\eqref{eq:Esp}, associated with the OAM photon with the momentum $m'_s$ emerging from the end of the waveguide at $z=L$
where the envelopes $a_{m_s\rightarrow m_s'}(z)$ and $a_{m'_s\rightarrow m_s'}(z)$ are solutions to the coupled mode equations obtained from Eq.~\eqref{CME} by replacing
the boundary conditions with $a_{m_s\rightarrow m_s'}(z=L) = 0$ and $a_{m_s'\rightarrow m_s'} (z=L) = 1$:
\begin{subequations}
\begin{align}
    a_{m_s\rightarrow m_s'}(z) &= -i\frac{\kappa_s}{\gamma_s}e^{-i [d_s (L+z)-\delta_s (L/v_g^{m'_s}-z/v_g^{m_s})]} \\ \nn &\ \ \ \times\sin[\gamma_s(L-z)],\\
     a_{m'_s\rightarrow m_s'}(z) &= e^{i (L-z)(-d_s+ \delta_s/v_g^{m'_s})}\\ \nn &\ \ \ \times \left\{\cos[\gamma_s(L-z)]+i\frac{d_s}{\gamma_s}\sin[\gamma_s(L-z)]\right\}.
\end{align}
\label{eq:as2}
\end{subequations}


\subsection{Idler photons}

We can similarly define the asymptotic-out modes of the left-propagating \textit{idler} photons, given in Eqs.~\eqref{eq:Ei}and~\eqref{eq:Eip}. In our convention dictates that, we assume that group velocities, as well as the wavenumbers are always positive, and the direction of propagation of the idler photons (towards $z\rightarrow -\infty$) is explicitly noted in the ansatz for the field (see discussion following Eq.~\eqref{eq:E}).


The asymptotic-out idler modes associated with $\mathbf{O}_{-m_i}^{\sigma_i}$ photons emerging from the $z=0$ end of the waveguide are given by Eq.~\ref{eq:Ei}, with envelopes fulfilling boundary conditions $a_{-m_i\leftarrow -m_i}(z=0)=1$ and $a_{-m_i\leftarrow -m_i'}(z=0)=0$, grating-coupled mode equations
\begin{subequations}
\begin{align}
    -&i\frac{\text{d}}{\text{d}z} a_{-m_i\leftarrow -m_i}(z)
    +\frac{\omega_s-\omega_{t,i}}{v_g^{m_i}}\; a_{-m_i\leftarrow -m_i}(z) \\
    &+ \kappa_s \; a_{-m_i\leftarrow -m_s'}(z) \;e^{-i\qty[ k_i^{m'_i}(\omega_i) -k_i^{m_i}(\omega_i)-K_i]z} =0, \nn\\
    -&i\frac{\text{d}}{\text{d}z} a_{-m_i\leftarrow -m_i'}(z)
    +\frac{\omega_i-\omega_{t,i}}{v_g^{m'_i}}\; a_{-m_i\leftarrow -m_i'}(z) \\
    &+ \kappa_i^* \;a_{-m_i\leftarrow -m_i}(z) \;e^{i\qty[k_i^{m'_i}(\omega_i) -k_i^{m_i}(\omega_i)-K_i]z} =0,\nn
\end{align}\label{coupled.modes.Idler}
\end{subequations}
derived by swapping the sign on the spatial coordinate $z$ in Eqs.~\eqref{coupled.modes.HE}, to describe coupling between left-propagating modes. Consistently with the introduced convention, the wavenumbers $k_i$ and $k_i'$, as well as the group velocities $v_g^{m_i}$ and $v_g^{m_i'}$, are chosen as positive. The solutions are
\begin{subequations}
\begin{align}
    a_{-m_i\leftarrow -m_i}(z)
    &= e^{iz\qty(d_i+\delta_i/v_g^{m_i})}\\ \nn &\ \ \ \times \left\{\cos(\gamma_iz)-i\frac{d_i}{\gamma_i}\sin\qty(\gamma_iz)\right\},\\
    a_{-m_i\leftarrow -m_i'}(z)
    &= -i\,\frac{\kappa_i^*}{\gamma_i}\,e^{-i z\qty(d_s-\delta_i/v_g^{m'_i})} \sin(\gamma_i z),
\end{align}
\label{eq:ai1}
\end{subequations}
where $\delta_i=\omega_{t,i}-\omega_i$, $2d_i = \delta_i\qty(1/v_g^{-m_i}-1/v_g^{-m'_i}) + K_i+k_i^{-m_i}(\omega_i)-k_i^{-m'_i}(\omega_i)$, and $\gamma_i=\sqrt{d_i^2+|\kappa_i|^2}$.

Similarly, we express the asymptotic-out idler modes associated with $\mathbf{O}_{-m_i'}^{\sigma_i}$ photons emerging from the $z=0$ end of the waveguide, given by Eq.~\ref{eq:Eip} at with the boundary conditions $a_{-m_i'\leftarrow -m_i}(z=0) = 0$ and $a_{-m'_i\leftarrow -m_i'}(z=0) = 1$, with
\begin{subequations}
\begin{align}
    a_{-m_i'\leftarrow -m_i}(z) &= -i\frac{\kappa_i}{\gamma_i}e^{i z(d_i +\delta_i/v_g^{m_i})s} \sin(\gamma_iz),\\
     a_{-m_i'\leftarrow -m_i'}(z) &= e^{-iz (d_i- \delta_i/v_g^{m'_i})}\\ \nn &\ \ \ \times \left\{\cos(\gamma_i z)+i\frac{d_i}{\gamma_i}\sin(\gamma_iz)\right\}.
\end{align}
\label{eq:as2b}
\end{subequations}


\section{Generating pairs of OAM photons}
\label{sec:OAM_generation}

Below we list the explicit expressions for JSAs introduced in Eq.~\eqref{JSA_total}, complementing the expression for $\Phi^{m_s',\sigma_s;-m_i',\sigma_i}$ given in Eq.~\eqref{jsa.oam_approx}: 
\begin{align}\label{jsa.oam3}
 &\Phi^{m_s,\sigma_s;-m_i,\sigma_i}(\omega_s,\omega_i) \nn\\
     &=
      I^{m_s,\sigma_s;-m_i,\sigma_i}\,
     \int_0^L \mathrm{d}z \ e^{i\left(k_1-k_2- k_s^{m_s}
      + k_i^{-m_i} \right)z}\nn\\
      &\quad\times\left[a_{m_s\rightarrow m_s} (z) \right]^*
      \left[a_{-m_i \leftarrow -m_i} (z)\right]^*,
\end{align}
\begin{align}\label{jsa.oam4}
 &\Phi^{m_s,\sigma_s;-m_i',\sigma_i}(\omega_s,\omega_i) \nn\\
     &=
      I^{m_s,\sigma_s;-m_i,\sigma_i}\,
     \int_0^L \mathrm{d}z \ e^{i\left(k_1-k_2- k_s^{m_s}
      + k_i^{-m_i} \right)z}\nn\\
      &\quad\times\left[a_{m_s\rightarrow m_s} (z) \right]^*
      \left[a_{-m_i' \leftarrow -m_i} (z)\right]^*,
\end{align}
\begin{align}\label{jsa.oam5}
 &\Phi^{m_s',\sigma_s;-m_i,\sigma_i}(\omega_s,\omega_i) \nn\\
     &=
      I^{m_s,\sigma_s;-m_i,\sigma_i}\,
     \int_0^L \mathrm{d}z \ e^{i\left(k_1-k_2- k_s^{m_s}
      + k_i^{-m_i} \right)z}\nn\\
      &\quad\times\left[a_{m_s\rightarrow m_s'} (z) \right]^*
      \left[a_{-m_i \leftarrow -m_i} (z)\right]^*.
\end{align}

\bibliography{bibliography}

\end{document}